\documentclass[reprint,prl,aps,superscriptaddress]{revtex4-2}
\usepackage{amsmath,amssymb,amsfonts}
\usepackage{graphicx}
\usepackage{booktabs}
\usepackage{hyperref}
\usepackage{tikz,xcolor}
\usepackage[normalem]{ulem}

\definecolor{lime}{HTML}{A6CE39}
\DeclareRobustCommand{\orcidicon}{
	\begin{tikzpicture}
	\draw[lime, fill=lime] (0,0) 
	circle [radius=0.16] 
	node[white] {{\fontfamily{qag}\selectfont \tiny ID}};
	\draw[white, fill=white] (-0.0625,0.095) 
	circle [radius=0.007];
	\end{tikzpicture}
	\hspace{-2mm}
}
\foreach \x in {A, ..., Z}{%
	\expandafter\xdef\csname orcid\x\endcsname{\noexpand\href{https://orcid.org/\csname orcidauthor\x\endcsname}{\noexpand\orcidicon}}
}

\begin{document}

\title{Persistent State Method for Resonances on Classical and Quantum Computers}

\author{Cong-Wu Wang\orcidA{}}
\affiliation{Key Laboratory of Nuclear Physics and Ion-beam Application (MOE), Institute of Modern Physics, Fudan University, Shanghai 200433, China}
\affiliation{Ruhr-Universit\"at Bochum, Fakult\"at f\"ur Physik und Astronomie, Institut f\"ur Theoretische Physik II, D-44780 Bochum, Germany}
\affiliation{Shanghai Research Center for Theoretical Nuclear Physics, NSFC and Fudan University, Shanghai 200438, China}

\author{Lukas Bovermann\orcidB{}}
\affiliation{Ruhr-Universit\"at Bochum, Fakult\"at f\"ur Physik und Astronomie, Institut f\"ur Theoretische Physik II, D-44780 Bochum, Germany}

\author{Evgeny Epelbaum\orcidC{}}
\affiliation{Ruhr-Universit\"at Bochum, Fakult\"at f\"ur Physik und Astronomie, Institut f\"ur Theoretische Physik II, D-44780 Bochum, Germany}

\author{Hermann Krebs\orcidD{}}
\affiliation{Ruhr-Universit\"at Bochum, Fakult\"at f\"ur Physik und Astronomie, Institut f\"ur Theoretische Physik II, D-44780 Bochum, Germany}

\author{Dean Lee\orcidE{}}
\affiliation{Facility for Rare Isotope Beams and Department of Physics and Astronomy, Michigan State University, East Lansing, Michigan 48824, USA}

\author{Yu-Gang Ma\orcidF{}}
\affiliation{Key Laboratory of Nuclear Physics and Ion-beam Application (MOE), Institute of Modern Physics, Fudan University, Shanghai 200433, China}
\affiliation{Shanghai Research Center for Theoretical Nuclear Physics, NSFC and Fudan University, Shanghai 200438, China}
\affiliation{School of Physics, East China Normal University, Shanghai 200241, China}

\author{Avik Sarkar\orcidG{}}
\affiliation{Department of Physics, Tohoku University, Aoba 6-3, Sendai, 980-8578, Miyagi, Japan}
\affiliation{Institute for Advanced Simulation (IAS-4), Forschungszentrum J\"ulich, D-52425 J\"ulich, Germany}

\begin{abstract}
We introduce a new method for extracting resonance pole positions that does not require non-Hermitian extensions of a Hamiltonian. The complex resonance poles are extracted from unitary time evolution of a persistent state, a compact trial state chosen such that its survival amplitude is governed by a single exponential over an extended time window. We solve several two- and three-body systems interacting via short- and long-range forces on a lattice and show that the pole positions obtained using the persistent state method converge approximately exponentially with the linear size of the system. We also consider a gate-based quantum implementation of our method using the Rodeo algorithm and a Hamiltonian variational ansatz, and outline an extension to two-cluster scattering.
\end{abstract}

\maketitle

\textit{Introduction.}---Resonances occur throughout nuclear, hadronic, atomic, and molecular physics. A bound state is a normalizable eigenstate of a Hamiltonian, while resonances are associated with poles of the scattering matrix in the complex energy plane. The real part of the pole position gives the resonance energy, while the imaginary part is minus half the width. However, in a finite box, the continuous spectrum becomes discrete, and outgoing waves eventually reflect from the boundary. Here, we show that resonance poles can, nevertheless, be extracted with high accuracy directly from unitary time evolution of a system in a finite volume by choosing a proper compact trial state. Our benchmark calculations reveal an approximately exponential decrease in the resulting extraction error with the spatial extent of the finite volume.

Decays of resonant states can be naturally described using the survival or Loschmidt amplitude, the complex-valued probability amplitude for a quantum system to remain in a considered initial state. The spectral representation of the survival amplitude has been worked out by Fock and Krylov \cite{FockKrylov:1947}. For Hamiltonians bounded from below, decays of resonant states cannot remain exactly exponential at arbitrarily late times \cite{Khalfin:1958}. Non-exponential decays have been studied in nuclear \cite{PeshkinVolyaZelevinsky:2014,VolyaZelevinsky:2020,WangNazarewicz:2021,WangNazarewiczVolyaMa:2023} as well as in atomic and molecular systems \cite{NicolaidesMercouris:1996,Wilkinson:1997,Rothe:2006}.

Deviations from exponential decays generally originate from a non-resonant background. This raises a natural question: How strongly can such a background be suppressed by choosing the initial state? The central idea of the persistent state method (PSM) consists in optimizing a compact initial state in such a way that over a considered time window, its survival amplitude is dominated by a single resonance pole. For a constructed family of normalizable states, we show that the background-to-pole ratio can decrease exponentially, with a rate controlled by the nearest competing nonanalytic structure. For strictly compact trial states of spatial extent $L$, our numerical results in the considered range of $L$ are consistent with the error in the pole extraction scaling as ${\rm poly}(L)e^{-cL}$. Thus, PSM does not eliminate the eventual late-time non-exponential decay, but it can systematically suppress its effect over a controlled intermediate time window.

Several established techniques are available for extracting resonance pole positions. In particular, L{\"u}scher-type methods relate discrete finite-volume energies to infinite-volume scattering information \cite{Luscher:1986pf,Luscher:1990ux,Luscher:1991cf} and have been extended to resonances, coupled channels, and few-body systems \cite{Bernard:2008ax,Hansen:2012tf,Hansen:2014eka,Briceno:2014oea,Briceno:2017tce,Konig:2020lzo,Mai:2021lwb}. While these methods are powerful and systematic, resonance poles are obtained indirectly by first extracting scattering information from finite-volume energy levels, then parametrizing the amplitude and finally performing analytic continuation to complex energy. The accuracy of the pole extraction, therefore, depends on the precision and density of the finite-volume energy spectrum. Likewise, stabilization and density-of-states methods infer resonance properties from how discrete levels change with the box or basis size \cite{HaziTaylor:1970,SlimakJordanFalcetta:2021}.

Complex scaling methods \cite{Aguilar:1971ve,Ho:1983zz,Moiseyev:1998zz} allow one to determine resonance locations directly as discrete complex eigenvalues and can provide accurate results for pole positions. This is achieved by analytically continuing the coordinates or Hamiltonian in the complex plane. Berggren-basis approaches, including the Gamow shell model \cite{Berggren:1968zz,IdBetan:2002kca,Michel:2008rt}, take a different route: they expand the state in bound, resonant, and nonresonant continuum states defined along a contour in the complex-energy plane, leading after discretization to a complex-symmetric eigenvalue problem. Realistic interactions can be treated through suitable basis transformations. 
Recent work has combined eigenvector continuation with complex scaling and
Berggren-basis calculations, extending these ideas to broad few- and many-body
resonances
\cite{Yapa:2023xyf,Yu:2023ucq,Yapa:2024lya,Zhang:2024gac,Zhang:2024ril}.
Analytic continuation in the coupling constant has also recently been
implemented within nuclear lattice effective field theory to extract the
$^5$He resonance energy and width from auxiliary-field quantum Monte Carlo
calculations \cite{Niu:2026resonance}.
These methods are powerful, but high accuracy usually requires a controlled
analytic continuation, transformation, or extrapolation away from the original
Hermitian Hamiltonian. If the interaction is only available numerically, as is common in lattice calculations \cite{Elhatisari:2022zrb}, this step represents a highly nontrivial task that can lead to large numerical errors.

Time-dependent wave-packet methods aim to extract resonance energies and lifetimes from real-time decay of a generic normalizable initial state \cite{AguilarGalindoBorisovDiazTendero:2021}. The PSM proposed here can be viewed as a higher-accuracy extension of these techniques to finite volume. Instead of evolving a generic wave packet, we optimize a compact initial state such that its survival amplitude is dominated by the resonance pole as much as possible over an extended time window.

In this paper, we aim for a high-accuracy pole extraction from a finite-volume Hamiltonian available in a numerical form and, therefore, use the complex absorbing potential (CAP) method \cite{Riss:1993,Muga:2004} for benchmarking our results. This approach can be applied to finite-volume Hamiltonians and does not require performing analytic continuation of the interaction. Resonance poles are obtained by adding an imaginary absorber near the outer boundary and solving the resulting eigenvalue problem. This makes the CAP method well suited to high-accuracy finite-volume calculations even in situations where the interaction is only available numerically. This is, however, achieved at the cost of making the Hamiltonian non-Hermitian. It is also worth mentioning that the absorbing potential in the CAP method needs to be tuned: If the absorber is too weak, waves reach the wall; if it increases too steeply, the absorber itself causes reflection. Both the PSM and CAP methods can, therefore, be used to 
extract resonance pole positions for finite-volume Hamiltonians available in a numerical form. In the CAP approach, this is achieved by changing the Hamiltonian, whereas the PSM only changes the initial state.

This difference matters especially for quantum computing. Quantum gates naturally implement unitary dynamics, so time evolution within the CAP method usually needs a larger unitary construction, such as dilation or block-encoding, together with ancillas, measurements, and possibly postselection. The PSM instead keeps the physical Hamiltonian Hermitian and uses ordinary unitary time evolution. It has two ingredients: \textit{compactness}, which keeps the initial state away from the outer boundary, and \textit{persistence}, which makes the survival amplitude nearly a single resonance exponential over a chosen time window. The pole is then extracted directly from the unitary dynamics.

\textit{Persistent State Method.}---For an isolated resonance, we start with a
compact single-cluster state $|\Psi_1\rangle$ that vanishes outside a radius
$L_{\text{init}}$,
\[
\langle \mathbf r | \Psi_1 \rangle \equiv \Psi_1(\mathbf r)=0,\qquad r>L_{\text{init}}.
\]
We choose $L_{\text{init}}$ well inside a larger 
volume
$L$ and evolve the state with the full Hermitian Hamiltonian $H$.
The survival, or Loschmidt, amplitude is defined as
\begin{equation}
  A_L(t)=\langle\Psi_1|e^{-iHt}|\Psi_1\rangle . \label{eq:survival}
\end{equation}
The idea of persistence is easiest to see in the corresponding infinite-volume system. After
analytic continuation, the survival amplitude separates into a
resonance-pole term and a nonresonant background,
\[
A_\infty(t)=\mathcal R_R e^{-iz_Rt}+I_{\text{bg}}(t),
\qquad
z_R=E_R-i\Gamma/2 ,
\]
as discussed in Supplemental Material \cite{SM}, Sec.~\ref{sec:supp-contour}. We tune the compact
initial trial state such that its finite-volume survival amplitude over an intermediate time window 
is, as much as possible, dominated by a single exponential. 
We define the local complex energy via
\begin{equation}
  E_{\text{loc}}(t)
  =\frac{i}{\tau}
  \ln\left[\frac{A_L(t+\tau)}{A_L(t)}\right].
\end{equation}
For a single-exponential form of the survival amplitude, $E_{\text{loc}}(t)=z_R$ is constant. We, therefore,
optimize the compact initial state by enforcing the real and imaginary parts of
$E_{\text{loc}}(t)$ to stay as constant as possible, see Sec.~\ref{sec:supp-psm-implementation} of \cite{SM} for details. The plateau gives the resonance energy and width.

The appropriate fitting window starts after the short-time region and ends before a
significant boundary reflection can return. A simple one-dimensional (1D) nearest-neighbor
lattice hopping example shows why this suppresses finite-volume effects.  For this case, we have the following results for the dispersion relation, group velocity, and maximum group velocity:
\begin{displaymath}
 E(k)=\frac{1-\cos(ka)}{ma^2},\quad
 v_g(k)=\frac{\sin(ka)}{ma},\quad
 v_{\max}=\frac{1}{ma},
\end{displaymath}
where $m$ is the mass and $a$ is the lattice spacing. If a round trip of
distance $D$ must occur in a time $t$ with $D/t>v_{\max}$, no real lattice
mode can make the trip. The remaining propagation is evanescent-like and is
exponentially suppressed with distance. If $D$ and $t$ both grow in
proportion to the linear system size $L$, while $D/t>v_{\max}$ remains fixed,
the suppression has the form $\exp(-cL)$ with $c>0$ (see Ref.~\cite{SM},
Sec.~\ref{sec:supp-psm-scaling}). Thus, if the fitting window stays a fixed fraction shorter than the fastest
round-trip time, boundary-return effects become exponentially small as $L$
grows. Other lattices have different dispersion relations and velocity scales.

\textit{Examples and benchmarks.}---We first study convergence of our method with linear system size using a 1D short-range two-body lattice model in the normalized even-parity half-line basis.  The interaction has only two nonzero matrix elements,
\begin{equation}
 V_n=-4\delta_{n0}+3\delta_{n2},
 \label{eq:main-short-range-model}
\end{equation}
with nearest-neighbor kinetic energy and $t_{\rm hop}=1/2$.  This simple model is useful because its resonance pole can be calculated to very high accuracy.  For each $L$, PSM varies the initial state only inside
$L_{\rm init}=0.2L$ and uses the time window
$t\in[0.9L,1.1L]$. The upper endpoint lies below the fastest boundary return; see Ref.~\cite{SM}, Sec.~\ref{sec:supp-psm-scaling}.  The state is optimized without using the actual pole position.

\begin{figure}[t]
  \centering
  \includegraphics[width=\columnwidth]{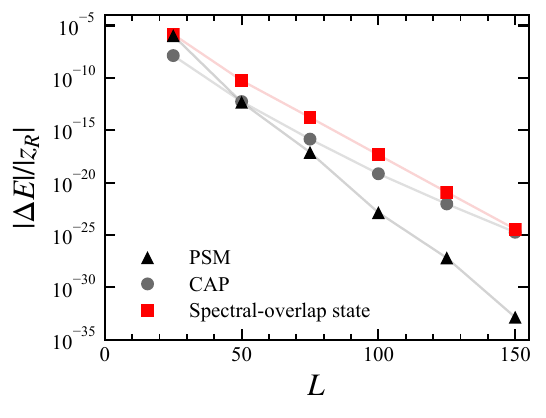}
  \caption{\label{fig:psm-mp-comparison}
  Relative resonance-pole error versus linear system size for the PSM, CAP, and the pole-informed spectral-overlap state in the 1D short-range lattice model. In the PSM, the state is varied only inside $L_{\rm init}=0.2L$.  Its error decreases approximately exponentially with $L$ over the displayed range, while the optimized CAP error follows the stretched-exponential form $\exp(-cL^{2/3})$. }
\end{figure}

Figure~\ref{fig:psm-mp-comparison} shows the relative error of the extracted resonance pole as $L$ increases. The PSM points are nearly linear on the semilog plot, showing an approximately exponential decrease with $L$ over the considered range,
\begin{equation}
 \frac{|E_{\rm PSM}-z_R|}{|z_R|}
 \sim {\rm poly}(L)\,e^{-c_{\rm PSM}L},
 \qquad c_{\rm PSM}>0 ,
 \label{eq:main-psm-exp-convergence}
\end{equation}
and reaching $\sim 1.4\times10^{-33}$ at $L=150$. Here and below, exponential finite-size convergence refers to the linear extent $L$, not the spatial volume $L^d$.
For the optimized CAP calculation, a slower stretched-exponential convergence behavior of the form $\exp(-c_{\rm CAP}L^{2/3})$ is found (see Sec.~\ref{sec:supp-cap-scaling} of Ref.~\cite{SM}).

The spectral-overlap curve in Fig.~\ref{fig:psm-mp-comparison} is a pole-informed benchmark, not a method for determining an unknown pole. For each $L$, the known $z_R$ is used to construct a normalizable state whose Breit--Wigner spectral density is multiplied by a weighting function that suppresses the background arising from the continuum threshold and the two lattice band edges. This state is evolved with the same Hermitian Hamiltonian as in PSM, and its survival amplitude is fitted over the fixed-width window $t\in[L-6.7,L+6.7]$, without state optimization. The fitted pole converges exponentially to the input $z_R$ as $L$ increases, following the predicted behavior $L^{-1/2}e^{-c_{\rm SO}L}$ where $c_{\rm SO}$ is a constant, as proven in Sec.~\ref{sec:supp-contour} of Ref.~\cite{SM}.

For PSM, the optimized spectral overlap is not known analytically, so its full exponential convergence is shown numerically here rather than proven in general. We prove for the 1D nearest-neighbor example that the 
effects from boundary reflection are exponentially suppressed 
when the fitting window ends before the fastest real-momentum round trip.

\begin{figure*}[t!]
  \centering
  \includegraphics[width=\textwidth]{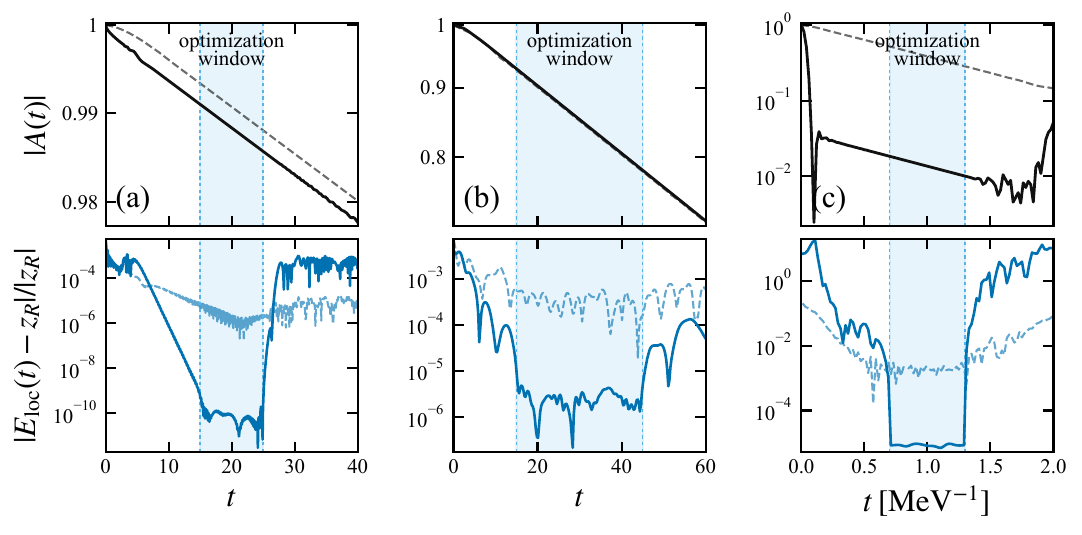}
 \vskip -0.4 true cm
  \caption{\label{fig:surv_amps}
  Application of the PSM to (a) a 1D two-body lattice system with an unscreened Coulomb tail, (b) the corresponding central interaction on a 3D cubic lattice, and (c) a 1D three-body lattice system.  Upper panels show the survival amplitude $|A(t)|$ and lower panels show the relative local-energy error $|E_{\rm loc}(t)-z_R|/|z_R|$.  Solid (dashed) curves correspond to the optimized (original) initial states, while shaded regions mark the time windows used for optimization.  The reference pole positions $z_R$ are obtained from solving the lattice Schr\"odinger equation in (a) and large-volume CAP calculations in (b) and (c).}
\end{figure*}

We next test the applicability of the PSM to systems with long-range interactions. We start with a 1D radial two-body problem on a half-line and choose the interaction to consist of a short-range Gaussian well and barrier, together with an unscreened repulsive Coulomb tail,
\begin{equation}   
V(r)=-8e^{-0.16r^2}+4e^{-0.04r^2}+\frac{1}{r}, 
 \label{eq:main-coulomb-model}
\end{equation}
with $r=na$, the lattice spacing $a=0.25$ and the reduced mass $\mu=1$.
Using $L=60$ and $L_{\rm init}=7$, we optimize the initial
state over the time window $t\in[15,25]$, as detailed in Sec.~\ref{sec:supp-psm-implementation} of Ref.~\cite{SM}, see Fig.~\ref{fig:surv_amps}(a), the resonance position is found to be 
\begin{equation}
E_{\rm PSM}=1.081754052720-0.000536238798\,i .
\end{equation}
This result agrees with the reference value of $z_R=1.081754052696-0.000536238776 \,i$, obtained by solving a discretized version of the Schr\"odinger equation in infinite volume (see Sec.~\ref{sec:supp-benchmarks} of Ref.~\cite{SM}), to a relative accuracy $|\Delta E|/|z_{R}| \sim 3.0\times10^{-11}$.

Our second example places the same interaction potential on a 3D cubic lattice with spacing $a=1$.  The Coulomb interaction at the origin is regularized as $1/(r+a/2)$, and the calculation is reduced to radial states in the $A_1$ irreducible representation of the cubic group. We use $L=60$, $L_{\rm init}=7$, and the time window of $t\in[15,45]$, see Fig.~\ref{fig:surv_amps}(b). The PSM result,
\begin{equation}
E_{\rm PSM}=1.3826301-0.0057088\,i ,
\end{equation}
agrees with the large-volume CAP reference
$z_R=1.3826296-0.0057115\,i$ with a relative error of $|\Delta E|/|z_{R}| \sim 2.0\times10^{-6}$.

As a final example, we consider a system of three equal-mass particles of mass $m=938.92\,{\rm MeV} $ in one dimension. 
Using Jacobi coordinates $y_1=x_2-x_1$ and
$y_2=x_3-(x_1+x_2)/2$, where $x_i$ are the particle positions,
the repulsive two-body and attractive three-body Gaussian interactions are
\begin{eqnarray}
 V_{\rm 2B}&=&V_+ \, \big[
 e^{-y_1^2/R_+^2}
 +e^{-(y_2-y_1/2)^2/R_+^2}
 +e^{-(y_2+y_1/2)^2/R_+^2}\big],\nonumber\\
 V_{\rm 3B}&=&-V_-\, 
 e^{-[y_1^2+(y_2-y_1/2)^2+(y_2+y_1/2)^2]/R_-^2},
 \label{eq:main-threebody-model}
\end{eqnarray}
where
$V_+=30\,{\rm MeV}$, $R_+=1\,{\rm fm}$,
$V_-=100\,{\rm MeV}$, $R_-=2\,{\rm fm}$, and lattice spacing $a=1\,{\rm fm}$.
Using $L=60\,{\rm fm}$, $L_{\rm init}=4\,{\rm fm}$, and the optimization window 
$t\in[0.7,1.3]\,{\rm MeV}^{-1}$ as shown in Fig.~\ref{fig:surv_amps}(c), the extracted resonance position is found to be 
\begin{equation}
E_{\rm PSM}=8.1847589 - 1.0135397\,i\ {\rm MeV}.
\end{equation}
This result agrees with the reference value of $z_R=8.1846910-1.0135355\,i\ {\rm MeV}$, obtained by performing a large-volume CAP calculation as described in Sec.~\ref{sec:supp-benchmarks} of Ref.~\cite{SM} with a relative error of $|\Delta E|/|z_{R}| \sim 8.3\times10^{-6}$.

To summarize, we considered several examples featuring different systems, Hamiltonians, and resonance decay widths. In every case, optimizing the compact state makes the survival amplitude nearly exponential over the chosen time window as visualized in Fig.~\ref{fig:surv_amps}. Without changing the form of the Hamiltonian, we are able to extract the resonance pole positions with relative errors of less than $10^{-5}$.

\begin{table*}[t!]
\caption{\label{tab:quantum_comp} Ideal noiseless simulation of the quantum PSM for the 1D two-body resonance. The independent lattice Schr\"odinger equation result,
$E_{\rm ref}=1.08175405-0.000536239i$, is used only after the PSM extraction to calculate the errors. State preparation is restricted to $L_{\rm init}=7$.}
\begin{ruledtabular}
\begin{tabular}{ccccc}
Rodeo Cycles & HVA Layers & $E_{\rm Quantum}$ & $\Delta \mathrm{Re}\,E$ & $\Delta \mathrm{Im}\,E$ \\
\colrule
\phantom{0}6 & 2 & $1.0817577-0.00053574i$ & $3.7\times10^{-6}$ & $5.0\times10^{-7}$ \\
\phantom{0}6 & 3 & $1.0817556-0.00053570i$ & $1.6\times10^{-6}$ & $5.3\times10^{-7}$ \\
\phantom{0}6 & 4 & $1.0817528-0.00053542i$ & $1.2\times10^{-6}$ & $8.2\times10^{-7}$ \\
\colrule
\phantom{0}8 & 2 & $1.0817537-0.00053644i$ & $3.3\times10^{-7}$ & $2.0\times10^{-7}$ \\
\phantom{0}8 & 3 & $1.0817537-0.00053643i$ & $3.4\times10^{-7}$ & $1.9\times10^{-7}$ \\
\phantom{0}8 & 4 & $1.0817537-0.00053645i$ & $3.3\times10^{-7}$ & $2.1\times10^{-7}$ \\
\colrule
10 & 2 & $1.0817540-0.00053620i$ & $8.1\times10^{-8}$ & $3.7\times10^{-8}$ \\
10 & 3 & $1.0817541-0.00053628i$ & $7.3\times10^{-8}$ & $4.5\times10^{-8}$ \\
10 & 4 & $1.0817541-0.00053633i$ & $3.6\times10^{-8}$ & $9.5\times10^{-8}$ \\
\end{tabular}
\end{ruledtabular}
\end{table*}

\textit{Quantum Computing Implementation.}---As explained in the Introduction, the PSM relies on unitary time evolution and only requires measurements of the survival amplitude. On a gate-based quantum computer, one ancilla can measure the real and imaginary parts of $\langle\Psi|e^{-iHt}|\Psi\rangle$ with controlled time evolution. We consider a 1D two-body system specified in Eq.~(\ref{eq:main-coulomb-model}) and prepare the state inside $L_{\rm init}=7$. Starting from a truncated Gaussian, we apply a phase-symmetric single-ancilla rodeo algorithm filter \cite{Choi:2020pdg,Qian:2021wya,Bee-Lindgren:2022nqb,Gomes:2024zln}; a successful cycle applies $\cos[(H_{\rm init}-E_{\rm target})t_k]$. We use $t_k=k^{1.25}$ \cite{Cohen:2023rhd,Stetcu:2022nhy,Patkowski:2026hag} and a coarse target $E_{\rm target}=1.082$, which can come from a coarse energy scan, and then refine the state with a Hamiltonian variational ansatz (HVA) \cite{Wecker:2015hva,Ho:2019hva,Wiersema:2020hva}.

Table~\ref{tab:quantum_comp} shows an ideal noiseless simulation with 6, 8, and 10 rodeo cycles and HVA depths $p=2,3,4$; the total postselection probability is about $10\%$. Exact statevectors and automatic differentiation are used only to emulate and optimize the circuit classically. On quantum hardware, the loss would be built from measured survival amplitudes and passed to a classical optimizer. The precise reference value for the resonance pole is used only afterward to calculate the error.

Increasing the number of rodeo cycles gives the largest improvement. At six cycles, the HVA still helps. At eight and ten cycles, the rodeo-filtered state is already close to the optimum, so the HVA depth matters little. With ten cycles, all $p=2$--4 results agree with the lattice Schr\"odinger equation pole at about $10^{-7}$ or better in each component. These are ideal noiseless results, not predictions for noisy hardware, but they show that PSM can provide an efficient and accurate route to resonance calculations on a quantum computer.

\textit{Summary and Outlook.}---We have introduced the persistent state method for extracting resonance locations, which keeps the Hamiltonian Hermitian and handles the finite-volume problem through the choice of initial state. It allows one to extract resonance poles from ordinary real-time evolution by optimizing a compact trial state and using a time window before finite-volume effects from boundary reflections become important. For short-range interactions, our results indicate that the PSM error decreases approximately exponentially with the linear system size. An analytic spectral construction demonstrates how exponential pole isolation can arise, and we have also shown 
that boundary-return effects arising from propagation outside the group-velocity cone are exponentially suppressed. 

The considered applications to two- and three-body systems using different numbers of spatial dimensions and interaction ranges demonstrate that our method remains accurate beyond the simplest short-range problem, while the gate-based implementation shows that its main ingredients are compatible with unitary quantum evolution. These results establish the PSM as a reliable, robust, and systematically improvable way to calculate resonances without changing the physical Hamiltonian. 

The PSM can also be extended to scattering and reactions. For two-cluster scattering, compactness around one center is no longer useful. Instead, one can work in a periodic box starting with two localized clusters, and optimize a state for one standing wave in the box. During the evolution, probability can flow into other open channels. It would be interesting to explore the application of channel-resolved time-dependent amplitudes to the determination of the inelasticity parameters. Another interesting extension concerns the application of the PSM to Euclidean-time calculations using quantum Monte Carlo methods.

\begin{acknowledgments}
{\it Acknowledgments:} We are grateful for helpful discussions with Witek Nazarewicz, Simin Wang, and the members of the Nuclear Lattice Effective Field Theory Collaboration. C.W.W.~acknowledges helpful discussions with Lu Meng. We acknowledge the following funding and computing resources: C.W.W.~and Y.G.M.~acknowledge the National Natural Science Foundation of China under Grants No.~12547102 and 12422509 as well as the STCSM under Grant No.~23590780100. L.B., E.E., and H.K.~acknowledge the support by the European Research Council (ERC) under the European Union’s Horizon 2020 research and innovation programme (grant agreement No.~885150), by the MKW NRW under the funding code NW21-024-A, and by BMBF through the ErUM-Data project DEMOS.
D.L.~acknowledges the U.S.~Department of Energy grants DE-SC0013365, DE-SC0023175, DE-SC0026198, DE-SC0023658; U.S.~National Science Foundation grant PHY-2310620; as well as the Oak Ridge Leadership Computing Facility computing resources through the INCITE award “Ab-initio nuclear structure and nuclear reactions” and Michigan State University's Institute for Cyber-Enabled Research and High-Performance Computing Center.
\end{acknowledgments}

\bibliographystyle{apsrev4-2}
\bibliography{References}

@misc{SM,
    author = {},
    note = {See Supplemental Material at [URL], for additional derivations and numerical details.}
}

@article{Choi:2020pdg,
    author = {Choi, Kenneth and Lee, Dean and Bonitati, Joey and Qian, Zhengrong and Watkins, Jacob},
    title = "{Rodeo Algorithm for Quantum Computing}",
    eprint = "2009.04092",
    archivePrefix = "arXiv",
    primaryClass = "quant-ph",
    doi = "10.1103/PhysRevLett.127.040505",
    journal = "Phys. Rev. Lett.",
    volume = "127",
    number = "4",
    pages = "040505",
    year = "2021"
}

@article{Moiseyev:1998zz,
    author = "Moiseyev, Nimrod",
    title = "{Quantum theory of resonances: calculating energies, widths and cross-sections by complex scaling}",
    doi = "10.1016/S0370-1573(98)00002-7",
    journal = "Phys. Rept.",
    volume = "302",
    pages = "212--293",
    year = "1998"
}

@article{Ho:1983zz,
    author = "Ho, Y. K.",
    title = "{The method of complex coordinate rotation and its applications to atomic collision processes}",
    doi = "10.1016/0370-1573(83)90112-6",
    journal = "Phys. Rept.",
    volume = "99",
    pages = "1--68",
    year = "1983"
}

@article{Aguilar:1971ve,
    author = "Aguilar, J. and Combes, J. M.",
    title = "{A Class of analytic perturbations for one-body Schrodinger Hamiltonians}",
    doi = "10.1007/BF01877510",
    journal = "Commun. Math. Phys.",
    volume = "22",
    pages = "269--279",
    year = "1971"
}

@article{Michel:2008rt,
    author = "Michel, N. and Nazarewicz, W. and Ploszajczak, M. and Vertse, T.",
    title = "{Shell model in the complex energy plane}",
    eprint = "0810.2728",
    archivePrefix = "arXiv",
    primaryClass = "nucl-th",
    doi = "10.1088/0954-3899/36/1/013101",
    journal = "J. Phys. G",
    volume = "36",
    pages = "013101",
    year = "2009"
}

@article{IdBetan:2002kca,
    author = "Id Betan, R. and Liotta, R. J. and Sandulescu, N. and Vertse, T.",
    title = {Two-Particle Resonant States in a Many-Body Mean Field},
    eprint = "nucl-th/0201077",
    archivePrefix = "arXiv",
    doi = "10.1103/PhysRevLett.89.042501",
    journal = "Phys. Rev. Lett.",
    volume = "89",
    pages = "042501",
    year = "2002"
}

@article{Berggren:1968zz,
    author = "Berggren, T.",
    title = "{On the use of resonant states in eigenfunction expansions of scattering and reaction amplitudes}",
    doi = "10.1016/0375-9474(68)90593-9",
    journal = "Nucl. Phys. A",
    volume = "109",
    pages = "265--287",
    year = "1968"
}

@article{Riss:1993,
    author = "Riss, U. V. and Meyer, H.-D.",
    title = "{Calculation of resonance energies and widths using the complex absorbing potential method}",
    doi = "10.1088/0953-4075/26/22/021",
    journal = "J. Phys. B: At. Mol. Opt. Phys.",
    volume = "26",
    pages = "4503",
    year = "1993"
}

@article{Muga:2004,
    author = "Muga, J. G. and Palao, J. P. and Navarro, B. and Egusquiza, I. L.",
    title = "{Complex absorbing potentials}",
    doi = "10.1016/j.physrep.2004.03.002",
    journal = "Phys. Rept.",
    volume = "395",
    pages = "357--426",
    year = "2004"
}

@article{Luscher:1990ux,
    author = "L{\"u}scher, M.",
    title = "{Two particle states on a torus and their relation to the scattering matrix}",
    reportNumber = "DESY-90-131",
    doi = "10.1016/0550-3213(91)90366-6",
    journal = "Nucl. Phys. B",
    volume = "354",
    pages = "531--578",
    year = "1991"
}

@article{Luscher:1986pf,
    author = "L{\"u}scher, M.",
    title = "{Volume Dependence of the Energy Spectrum in Massive Quantum Field Theories. 2. Scattering States}",
    reportNumber = "DESY-86-034",
    doi = "10.1007/BF01211097",
    journal = "Commun. Math. Phys.",
    volume = "105",
    pages = "153--188",
    year = "1986"
}

@article{Luscher:1991cf,
    author = "L{\"u}scher, Martin",
    title = "{Signatures of unstable particles in finite volume}",
    reportNumber = "DESY-91-052",
    doi = "10.1016/0550-3213(91)90584-K",
    journal = "Nucl. Phys. B",
    volume = "364",
    pages = "237--251",
    year = "1991"
}

@article{Bernard:2008ax,
    author = "Bernard, Veronique and Lage, Michael and Meissner, Ulf-G. and Rusetsky, Akaki",
    title = "{Resonance properties from the finite-volume energy spectrum}",
    eprint = "0806.4495",
    archivePrefix = "arXiv",
    primaryClass = "hep-lat",
    reportNumber = "HISKP-TH-08-08, FZJ-IKP(TH)-08-10",
    doi = "10.1088/1126-6708/2008/08/024",
    journal = "JHEP",
    volume = "08",
    pages = "024",
    year = "2008"
}

@article{Hansen:2012tf,
    author = "Hansen, Maxwell T. and Sharpe, Stephen R.",
    title = "{Multiple-channel generalization of Lellouch-L{\"u}scher formula}",
    eprint = "1204.0826",
    archivePrefix = "arXiv",
    primaryClass = "hep-lat",
    doi = "10.1103/PhysRevD.86.016007",
    journal = "Phys. Rev. D",
    volume = "86",
    pages = "016007",
    year = "2012"
}

@article{Briceno:2014oea,
    author = "Brice{\~n}o, Raul A.",
    title = "{Two-particle multichannel systems in a finite volume with arbitrary spin}",
    eprint = "1401.3312",
    archivePrefix = "arXiv",
    primaryClass = "hep-lat",
    reportNumber = "JLAB-THY-14-1833",
    doi = "10.1103/PhysRevD.89.074507",
    journal = "Phys. Rev. D",
    volume = "89",
    number = "7",
    pages = "074507",
    year = "2014"
}

@article{Briceno:2017tce,
    author = "Brice{\~n}o, Ra{\'u}l A. and Hansen, Maxwell T. and Sharpe, Stephen R.",
    title = "{Relating the finite-volume spectrum and the two-and-three-particle $S$ matrix for relativistic systems of identical scalar particles}",
    eprint = "1701.07465",
    archivePrefix = "arXiv",
    primaryClass = "hep-lat",
    reportNumber = "JLAB-THY-17-2400",
    doi = "10.1103/PhysRevD.95.074510",
    journal = "Phys. Rev. D",
    volume = "95",
    number = "7",
    pages = "074510",
    year = "2017"
}

@article{Hansen:2014eka,
    author = "Hansen, Maxwell T. and Sharpe, Stephen R.",
    title = "{Relativistic, model-independent, three-particle quantization condition}",
    eprint = "1408.5933",
    archivePrefix = "arXiv",
    primaryClass = "hep-lat",
    doi = "10.1103/PhysRevD.90.116003",
    journal = "Phys. Rev. D",
    volume = "90",
    number = "11",
    pages = "116003",
    year = "2014"
}

@article{Mai:2021lwb,
    author = {Mai, Maxim and D{\"o}ring, Michael and Rusetsky, Akaki},
    title = "{Multi-particle systems on the lattice and chiral extrapolations: a brief review}",
    eprint = "2103.00577",
    archivePrefix = "arXiv",
    primaryClass = "hep-lat",
    reportNumber = "JLAB-THY-21-3327",
    doi = "10.1140/epjs/s11734-021-00146-5",
    journal = "Eur. Phys. J. ST",
    volume = "230",
    number = "6",
    pages = "1623--1643",
    year = "2021"
}

@article{Konig:2020lzo,
    author = {K{\"o}nig, Sebastian},
    title = "{Few-body bound states and resonances in finite volume}",
    eprint = "2005.01478",
    archivePrefix = "arXiv",
    primaryClass = "hep-lat",
    doi = "10.1007/s00601-020-01550-8",
    journal = "Few Body Syst.",
    volume = "61",
    number = "3",
    pages = "20",
    year = "2020"
}

@article{Yapa:2023xyf,
    author = {Yapa, Nuwan and Fossez, K{\'e}vin and K{\"o}nig, Sebastian},
    title = "{Eigenvector continuation for emulating and extrapolating two-body resonances}",
    eprint = "2303.06139",
    archivePrefix = "arXiv",
    primaryClass = "nucl-th",
    doi = "10.1103/PhysRevC.107.064316",
    journal = "Phys. Rev. C",
    volume = "107",
    number = "6",
    pages = "064316",
    year = "2023"
}

@article{Yu:2023ucq,
    author = {Yu, Hang and Yapa, Nuwan and K{\"o}nig, Sebastian},
    title = "{Complex scaling in finite volume}",
    eprint = "2309.03196",
    archivePrefix = "arXiv",
    primaryClass = "nucl-th",
    doi = "10.1103/PhysRevC.109.014316",
    journal = "Phys. Rev. C",
    volume = "109",
    number = "1",
    pages = "014316",
    year = "2024"
}

@article{Yapa:2024lya,
    author = {Yapa, Nuwan and K{\"o}nig, Sebastian and Fossez, K{\'e}vin},
    title = "{Toward scalable bound-to-resonance extrapolations for few- and many-body systems}",
    eprint = "2409.03116",
    archivePrefix = "arXiv",
    primaryClass = "nucl-th",
    doi = "10.1103/PhysRevC.111.064318",
    journal = "Phys. Rev. C",
    volume = "111",
    number = "6",
    pages = "064318",
    year = "2025"
}

@article{Zhang:2024ril,
    author = "Zhang, Xilin",
    title = "{Non-Hermitian Quantum Mechanics Approach for Extracting and Emulating Continuum Physics Based on Bound-State-like Calculations}",
    eprint = "2408.03309",
    archivePrefix = "arXiv",
    primaryClass = "nucl-th",
    doi = "10.1103/5frj-w5xh",
    journal = "Phys. Rev. Lett.",
    volume = "135",
    number = "24",
    pages = "242501",
    year = "2025"
}

@article{Zhang:2024gac,
    author = "Zhang, Xilin",
    title = "{Non-Hermitian quantum mechanics approach for extracting and emulating continuum physics based on bound-state-like calculations: Detailed description}",
    eprint = "2411.06712",
    archivePrefix = "arXiv",
    primaryClass = "nucl-th",
    doi = "10.1103/4wbf-gzk5",
    journal = "Phys. Rev. C",
    volume = "112",
    number = "6",
    pages = "064605",
    year = "2025"
}

@article{PeshkinVolyaZelevinsky:2014,
    author = "Peshkin, Murray and Volya, Alexander and Zelevinsky, Vladimir",
    title = "{Non-exponential and oscillatory decays in quantum mechanics}",
    journal = "EPL",
    volume = "107",
    number = "4",
    pages = "40001",
    year = "2014",
    doi = "10.1209/0295-5075/107/40001"
}

@article{VolyaZelevinsky:2020,
    author = "Volya, Alexander and Zelevinsky, Vladimir",
    title = "{Time-dependent relaxation of observables in complex quantum systems}",
    journal = "J. Phys. Complex.",
    volume = "1",
    number = "2",
    pages = "025007",
    year = "2020",
    doi = "10.1088/2632-072X/ab79bc",
    eprint = "1905.11918",
    archivePrefix = "arXiv",
    primaryClass = "quant-ph"
}

@article{WangNazarewicz:2021,
    author = "Wang, S. M. and Nazarewicz, W.",
    title = "{Fermion Pair Dynamics in Open Quantum Systems}",
    journal = "Phys. Rev. Lett.",
    volume = "126",
    number = "14",
    pages = "142501",
    year = "2021",
    doi = "10.1103/PhysRevLett.126.142501",
    eprint = "2104.03195",
    archivePrefix = "arXiv",
    primaryClass = "nucl-th"
}

@article{WangNazarewiczVolyaMa:2023,
    author = "Wang, S. M. and Nazarewicz, W. and Volya, A. and Ma, Y. G.",
    title = "{Probing the nonexponential decay regime in open quantum systems}",
    journal = "Phys. Rev. Research",
    volume = "5",
    number = "2",
    pages = "023183",
    year = "2023",
    doi = "10.1103/PhysRevResearch.5.023183",
    eprint = "2211.11619",
    archivePrefix = "arXiv",
    primaryClass = "nucl-th"
}

@article{NicolaidesMercouris:1996,
    author = "Nicolaides, C. A. and Mercouris, T.",
    title = "{On the violation of the exponential decay law in atomic physics: ab initio calculation of the time-dependence of the He$^-$ $1s2p^2$ $^4P$ non-stationary state}",
    journal = "J. Phys. B: At. Mol. Opt. Phys.",
    volume = "29",
    number = "6",
    pages = "1151",
    year = "1996",
    doi = "10.1088/0953-4075/29/6/012"
}

@article{Wilkinson:1997,
    author = "Wilkinson, Steven R. and Bharucha, Cyrus F. and Fischer, Martin C. and Madison, Kirk W. and Morrow, Patrick R. and Niu, Qian and Sundaram, Bala and Raizen, Mark G.",
    title = "{Experimental evidence for non-exponential decay in quantum tunnelling}",
    journal = "Nature",
    volume = "387",
    pages = "575--577",
    year = "1997",
    doi = "10.1038/42418"
}

@article{Rothe:2006,
    author = "Rothe, C. and Hintschich, S. I. and Monkman, A. P.",
    title = "{Violation of the Exponential-Decay Law at Long Times}",
    journal = "Phys. Rev. Lett.",
    volume = "96",
    number = "16",
    pages = "163601",
    year = "2006",
    doi = "10.1103/PhysRevLett.96.163601"
}

@article{HaziTaylor:1970,
    author = "Hazi, A. U. and Taylor, H. S.",
    title = "{Stabilization Method of Calculating Resonance Energies: Model Problem}",
    journal = "Phys. Rev. A",
    volume = "1",
    number = "4",
    pages = "1109--1120",
    year = "1970",
    doi = "10.1103/PhysRevA.1.1109"
}

@article{SlimakJordanFalcetta:2021,
    author = "Slimak, Stephen R. and Jordan, Kenneth D. and Falcetta, Michael F.",
    title = "{Role of Overlap between the Discrete State and Pseudocontinuum States in Stabilization Calculations of Metastable States}",
    journal = "J. Phys. Chem. A",
    volume = "125",
    number = "20",
    pages = "4401--4408",
    year = "2021",
    doi = "10.1021/acs.jpca.1c02699"
}

@article{AguilarGalindoBorisovDiazTendero:2021,
    author = "Aguilar-Galindo, Fernando and Borisov, Andrey G. and D{\'\i}az-Tendero, Sergio",
    title = "{Ultrafast Dynamics of Electronic Resonances in Molecules Adsorbed on Metal Surfaces: A Wave Packet Propagation Approach}",
    journal = "J. Chem. Theory Comput.",
    volume = "17",
    number = "2",
    pages = "639--654",
    year = "2021",
    doi = "10.1021/acs.jctc.0c01031"
}

@article{FockKrylov:1947,
    author = "Fock, V. and Krylov, N. S.",
    title = "{On the uncertainty relation between time and energy}",
    journal = "J. Phys. (USSR)",
    volume = "11",
    number = "2",
    pages = "112--120",
    year = "1947"
}

@article{Khalfin:1958,
    author = "Khalfin, L. A.",
    title = "{Contribution to the Decay Theory of a Quasi-Stationary State}",
    journal = "Sov. Phys. JETP",
    volume = "6",
    number = "6",
    pages = "1053--1063",
    year = "1958"
}

@article{Winter:1961,
    author = "Winter, Rolf G.",
    title = "{Evolution of a Quasi-Stationary State}",
    journal = "Phys. Rev.",
    volume = "123",
    number = "4",
    pages = "1503--1507",
    year = "1961",
    doi = "10.1103/PhysRev.123.1503"
}

@article{Niu:2026resonance,
    author  = {Niu, Zhong-Wang and Zhang, Shi-Sheng and Lu, Bing-Nan},
    title   = {Extracting Resonance Width from Lattice Quantum Monte Carlo Simulations Using Analytical Continuation Method},
    journal = {Physics Letters B},
    volume = {880},
    pages   = {140823},
    year    = {2026},
    doi     = {10.1016/j.physletb.2026.140823},
    eprint  = {2603.25081},
    archivePrefix = {arXiv},
    primaryClass  = {nucl-th}
}

@article{Patkowski:2026hag,
    author = "Patkowski, Matthew and Ayyildiz, Onat and Hunt, Katherine and Jansen, Nathan and Lee, Dean",
    title = "{Improved Rodeo Algorithm Performance for Spectral Functions and State Preparation}",
    eprint = "2602.05978",
    archivePrefix = "arXiv",
    primaryClass = "quant-ph",
    month = "2",
    journal = "",
    year = "2026"
}

@article{Cohen:2023rhd,
    author = "Cohen, Thomas D. and Oh, Hyunwoo",
    title = "{Optimizing the rodeo projection algorithm}",
    eprint = "2305.19952",
    archivePrefix = "arXiv",
    primaryClass = "quant-ph",
    doi = "10.1103/PhysRevA.108.032422",
    journal = "Phys. Rev. A",
    volume = "108",
    number = "3",
    pages = "032422",
    year = "2023"
}

@article{Stetcu:2022nhy,
    author = "Stetcu, I. and Baroni, A. and Carlson, J.",
    title = "{Projection algorithm for state preparation on quantum computers}",
    eprint = "2211.10545",
    archivePrefix = "arXiv",
    primaryClass = "quant-ph",
    reportNumber = "LA-UR-22-30149 ver 2, LA-UR-22-30149, LA-UR-22-30149 ver 3",
    doi = "10.1103/PhysRevC.108.L031306",
    journal = "Phys. Rev. C",
    volume = "108",
    number = "3",
    pages = "L031306",
    year = "2023"
}

@article{Bee-Lindgren:2022nqb,
    author = "Bee-Lindgren, Max and Qian, Zhengrong and DeCross, Matthew and Brown, Natalie C. and Gilbreth, Christopher N. and Watkins, Jacob and Zhang, Xilin and Lee, Dean",
    title = "{Controlled gate networks: theory and application to eigenvalue estimation}",
    eprint = "2208.13557",
    archivePrefix = "arXiv",
    primaryClass = "quant-ph",
    doi = "10.1140/epja/s10050-025-01750-y",
    journal = "Eur. Phys. J. A",
    volume = "61",
    number = "11",
    pages = "263",
    year = "2025"
}

@article{Qian:2021wya,
    author = "Qian, Zhengrong and Watkins, Jacob and Given, Gabriel and Bonitati, Joey and Choi, Kenneth and Lee, Dean",
    title = "{Demonstration of the rodeo algorithm on a quantum computer}",
    eprint = "2110.07747",
    archivePrefix = "arXiv",
    primaryClass = "quant-ph",
    doi = "10.1140/epja/s10050-024-01373-9",
    journal = "Eur. Phys. J. A",
    volume = "60",
    number = "7",
    pages = "151",
    year = "2024"
}

@article{Gomes:2024zln,
    author = "Gomes, Raphael Fortes Infante and Rocha, Julio Cesar Siqueira and Nogueira, Wallon Anderson Tadaiesky and Dias, Rodrigo Alves",
    title = "{Unraveling the rodeo algorithm through the Zeeman model}",
    eprint = "2407.11301",
    archivePrefix = "arXiv",
    primaryClass = "quant-ph",
    doi = "10.1088/1402-4896/add58c",
    journal = "Phys. Scripta",
    volume = "100",
    number = "6",
    pages = "065119",
    year = "2025"
}

@article{Wecker:2015hva,
    author = "Wecker, Dave and Hastings, Matthew B. and Troyer, Matthias",
    title = "{Progress towards practical quantum variational algorithms}",
    eprint = "1507.08969",
    archivePrefix = "arXiv",
    primaryClass = "quant-ph",
    doi = "10.1103/PhysRevA.92.042303",
    journal = "Phys. Rev. A",
    volume = "92",
    number = "4",
    pages = "042303",
    year = "2015"
}

@article{Wiersema:2020hva,
    author = "Wiersema, Roeland and Zhou, Cunlu and de Sereville, Yvette and
              Carrasquilla, Juan Felipe and Kim, Yong Baek and Yuen, Henry",
    title = "{Exploring Entanglement and Optimization within the Hamiltonian Variational Ansatz}",
    eprint = "2008.02941",
    archivePrefix = "arXiv",
    primaryClass = "quant-ph",
    doi = "10.1103/PRXQuantum.1.020319",
    journal = "PRX Quantum",
    volume = "1",
    number = "2",
    pages = "020319",
    year = "2020"
}

@article{Ho:2019hva,
    author = "Ho, Wen Wei and Hsieh, Timothy H.",
    title = "{Efficient variational simulation of non-trivial quantum states}",
    eprint = "1803.00026",
    archivePrefix = "arXiv",
    doi = "10.21468/SciPostPhys.6.3.029",
    journal = "SciPost Phys.",
    volume = "6",
    number = "3",
    pages = "029",
    year = "2019",
    primaryClass = "quant-ph"
}

@article{Elhatisari:2022zrb,
    author = "Elhatisari, Serdar and others",
    title = "{Wavefunction matching for solving quantum many-body problems}",
    eprint = "2210.17488",
    archivePrefix = "arXiv",
    primaryClass = "nucl-th",
    doi = "10.1038/s41586-024-07422-z",
    journal = "Nature",
    volume = "630",
    number = "8015",
    pages = "59--63",
    year = "2024"
}

\clearpage

\onecolumngrid

\setcounter{equation}{0}
\setcounter{table}{0}
\setcounter{section}{0}
\renewcommand{\theequation}{S\arabic{equation}}
\renewcommand{\thetable}{S\arabic{table}}
\renewcommand{\theHequation}{S\arabic{equation}}
\renewcommand{\theHtable}{S\arabic{table}}
\section*{Supplemental Material}
\setcounter{subsection}{0}
\renewcommand{\thesubsection}{\Alph{subsection}}
\setcounter{secnumdepth}{3}

This Supplemental Material provides the background and technical details supporting the main text. In Sec.~\ref{sec:supp-contour}, we derive the pole--background decomposition of the survival amplitude and analyze the suppression of the background in the infinite-volume continuum and on a finite-volume lattice. Sec.~\ref{sec:supp-benchmarks} defines the benchmark models and describes the calculations of their reference resonance poles.
Next, Sec.~\ref{sec:supp-psm-scaling} addresses finite-size effects for compact PSM states and their extension to higher-dimensional and many-body systems. In Secs.~\ref{sec:supp-cap-scaling} and \ref{sec:supp-psm-broad}, we discuss finite-volume effects within the CAP method and the treatment of broad resonances with PSM, respectively. Finally, Sec.~\ref{sec:supp-psm-implementation} provides the implementation details of the PSM calculations.

\subsection{Pole and background in the survival amplitude}
\label{sec:supp-contour}

This section combines analytic arguments with numerical calculations to understand the role played by different contributions in the survival amplitude. We first decompose the amplitude into a resonance-pole term and a nonresonant background. Sec.~\ref{subsec:SpectralConstruction} introduces a normalizable spectral-overlap construction and shows how the nonresonant background induces an error in the fitted pole. In Sec.~\ref{subsec:supp-essential-filter}, we consider an explicit infinite-volume construction and show that, when its parameter $\alpha$ and the fitting time are scaled together, the background-to-pole ratio decreases as ${\rm poly}(\alpha)e^{-\mu\alpha}$. In Sec.~\ref{subsec:supp-finite-lattice}, we extend the analysis to a finite-volume lattice Hamiltonian. Poisson summation separates the spectral-background and finite-volume contributions, whose errors decrease as ${\rm poly}(L)e^{-\mu L}$ along the prescribed scaling trajectory. Numerical calculations confirm these rates and show that either contribution can dominate, depending on the fitting window.

For a normalized initial state $|\Psi_1\rangle$, the survival amplitude is
$A_\infty(t)=\langle\Psi_1|e^{-iHt}|\Psi_1\rangle$. Its energy dependence can
be expressed through the projected resolvent
\begin{equation}
  G_{\Psi_1}(z)=
  \langle\Psi_1|(z-H)^{-1}|\Psi_1\rangle .
\end{equation}
We set the continuum threshold to $E=0$ and focus on the continuum
contribution. Any contribution from discrete bound states can be added
separately to the nonresonant background.

The continuum produces a branch cut along the positive real-energy axis.
Analytically continuing $G_{\Psi_1}(z)$ through this cut to the second
Riemann sheet, with the outgoing boundary condition, exposes the resonance
poles. We assume that the continued resolvent has an isolated pole at
$z_R$ with a nonzero residue $\mathcal R_R$. The nonzero
residue means that this resonance contributes to the survival amplitude of
$|\Psi_1\rangle$.

Using a clockwise contour $\mathcal C$ around the physical cut, the survival
amplitude is
\begin{equation}
  A_\infty(t)
  =-\frac{1}{2\pi i}\int_{\mathcal C}
   e^{-izt}G_{\Psi_1}(z)\,dz .
\end{equation}
For $t>0$, deforming the contour into the lower half-plane picks up the
resonance pole and leaves a background integral,
\begin{equation}
  A_\infty(t)
  =\mathcal R_R e^{-iz_Rt}
   +\frac{1}{2\pi i}\int_{\mathcal C_{\rm bg}}
    e^{-izt}G_{\rm II}(z)\,dz
  \equiv P(t)+B(t),
\end{equation}
where $G_{\rm II}$ denotes the resolvent on the second Riemann sheet.
The following subsections analyze how the background-to-pole ratio
$|B(t)/P(t)|$ can be suppressed.

\subsubsection{Normalizable spectral-overlap state construction}
\label{subsec:SpectralConstruction}
We first introduce pole-informed normalizable spectral-overlap states and derive how the background-to-pole ratio controls the error in the fitted resonance pole. The suppression of this ratio is analyzed in the following two subsections.

Using energy-normalized continuum eigenstates,
$\langle E|E'\rangle=\delta(E-E')$, the spectral theorem gives the
Fock--Krylov representation of the survival amplitude
\cite{FockKrylov:1947},
\begin{equation}
 A(t)=\langle\Psi|e^{-iHt}|\Psi\rangle
 =\int_0^\infty \rho_\Psi(E)e^{-iEt}\,dE,
 \qquad
 \rho_\Psi(E)=|\langle E|\Psi\rangle|^2\geq0,
 \qquad
 \int_0^\infty\rho_\Psi(E)\,dE=1 .
 \label{eq:supp-spectral-theorem}
\end{equation}

Let $z_R=E_R-i\gamma$, with $\gamma=\Gamma/2>0$, denote the resonance pole. We introduce a family of spectral-overlap functions $c_F(E)\equiv\langle E|\Psi_F\rangle$ whose modulus squared is
\begin{equation}
 \rho_F(E)=|c_F(E)|^2
 =\frac{1}{\mathcal N_F}
 \frac{F(E)}{(E-E_R)^2+\gamma^2}\,\Theta(E),
 \label{eq:supp-spectral-density}
\end{equation}
where $\mathcal N_F$ ensures normalization. The spectral weighting function $F(E)$ is real and nonnegative on the physical spectrum. For the contour analysis below, $F(E)$ is assumed to admit analytic continuation through the region of contour deformation away from the spectral endpoints and to satisfy $F(z_R)\neq0$.

Since the survival amplitude depends only on $|c_F(E)|^2$, the phase of $c_F(E)$ is irrelevant here, and we take $c_F(E)=\sqrt{\rho_F(E)}$. The resulting normalizable state is
\begin{equation}
 |\Psi_F\rangle
 =\int_0^\infty \sqrt{\rho_F(E)}\,|E\rangle\,dE .
 \label{eq:supp-spectral-state}
\end{equation}

Because the physical spectrum is bounded from below, the survival amplitude cannot be exactly exponential at all times \cite{Khalfin:1958}. For the state $|\Psi_F\rangle$, the pole--background decomposition introduced above takes the form
\begin{equation}
 A_F(t)=P_F(t)+B_F(t),\qquad
 P_F(t)=\frac{\pi}{\gamma\mathcal N_F}F(z_R)e^{-iz_Rt},
 \label{eq:supp-pole-background}
\end{equation}
where $B_F(t)=A_F(t)-P_F(t)$ contains the nonresonant contribution.

The ratio $B_F/P_F$ controls the bias produced when the survival amplitude is fitted by the exponential form $A_{\rm fit}(t)=Z_{\rm fit}e^{-iE_{\rm fit}t}$. Using the values of the survival amplitude at times $t_j\in[t_{\min},t_{\max}]$, we perform a least-squares fit of its logarithm which gives
\begin{equation}
 E_{\rm fit}
 =i\frac{\sum_j(t_j-\bar t)\log A_F(t_j)}
     {\sum_j(t_j-\bar t)^2},
 \qquad
 \bar t=\frac{1}{N}\sum_jt_j.
 \label{eq:supp-log-fit}
\end{equation}
Substituting Eq.~\eqref{eq:supp-pole-background} yields
\begin{equation}
 E_{\rm fit}-z_R
 =i\frac{\displaystyle\sum_j(t_j-\bar t)
 \log\!\left[1+\frac{B_F(t_j)}{P_F(t_j)}\right]}
 {\displaystyle\sum_j(t_j-\bar t)^2}.
 \label{eq:supp-fit-bias}
\end{equation}
Therefore, suppressing $B_F/P_F$ reduces the errors in both the real and imaginary parts of the fitted pole. The following subsections analyze this suppression in infinite volume and for a finite-volume lattice Hamiltonian.

\subsubsection{Threshold suppression in the infinite-volume continuum}
\label{subsec:supp-essential-filter}

We now choose an explicit form of the spectral weighting function in Eq.~\eqref{eq:supp-spectral-density} and use it to estimate analytically the background-to-pole ratio and the resulting pole-fitting error. We then test these predictions in the continuous-space Winter model. 

For a continuum spectrum with threshold at $E=0$, we choose the spectral weighting function
\begin{equation}
 F_\alpha(E)=\exp\left(-\alpha\frac{E_R}{E}\right),
 \qquad \alpha>0 .
 \label{eq:supp-essential-filter}
\end{equation}
It approaches zero at the threshold together with all of its derivatives. Using Eq.~\eqref{eq:supp-spectral-density}, the normalized survival amplitude is
\begin{equation}
 A_\alpha(t)
 =\frac{1}{\mathcal N_\alpha}
 \int_{\mathcal C_0}
 \frac{e^{S(z)}}{(z-z_R)(z-z_R^*)}\,dz,
 \qquad
 S(z)=-\frac{\alpha E_R}{z}-itz,
 \qquad
 \mathcal C_0=[0,\infty).
 \label{eq:supp-essential-integral}
\end{equation}
For $t>0$, the contour $\mathcal C_0$ can be deformed clockwise into a contour $\Gamma_{\rm end}$ in the lower-right quadrant that passes below $z_R$. Near the origin, $\operatorname{Re}(1/z)>0$ along this contour, so $\exp(-\alpha E_R/z)$ remains suppressed, while at infinity $e^{-itz}$ causes the integrand to decay. The residue theorem then gives
\begin{equation}
 \begin{aligned}
 A_\alpha(t)&=P_\alpha(t)+B_\alpha(t),\\
 P_\alpha(t)&=
 \frac{\pi}{\gamma\mathcal N_\alpha}
 \exp\left(-\frac{\alpha E_R}{z_R}-iz_Rt\right),\\
 B_\alpha(t)&=
 \frac{1}{\mathcal N_\alpha}
 \int_{\Gamma_{\rm end}}
 \frac{e^{S(z)}}{(z-z_R)(z-z_R^*)}\,dz .
 \end{aligned}
 \label{eq:supp-essential-contour}
\end{equation}

We estimate the threshold background using the method of steepest descent. When the exponent is large, the contribution from a saddle point $z_s$ satisfying $S'(z_s)=0$ is
\begin{equation}
 \int_{\Gamma_s}g(z)e^{S(z)}\,dz
 \sim
 g(z_s)e^{S(z_s)}
 \sqrt{\frac{2\pi}{-S''(z_s)}},
 \label{eq:supp-steepest-rule}
\end{equation}
up to a phase determined by the orientation of the steepest-descent contour. For Eq.~\eqref{eq:supp-essential-integral}, the saddle point accessible from $\mathcal C_0$ is
\begin{equation}
 S'(z_s)=\frac{\alpha E_R}{z_s^2}-it=0,
 \qquad
 z_s=e^{-i\pi/4}\sqrt{\frac{\alpha E_R}{t}} .
 \label{eq:supp-essential-saddle}
\end{equation}
At this saddle,
\begin{equation}
 \operatorname{Re}S(z_s)=-\sqrt{2\alpha E_Rt},
 \qquad
 |S''(z_s)|=\frac{2t^{3/2}}{(\alpha E_R)^{1/2}} .
\end{equation}
Substitution into Eq.~\eqref{eq:supp-steepest-rule} gives
\begin{equation}
 |B_\alpha(t)|
 \sim
 \frac{\sqrt{\pi}}{\mathcal N_\alpha}
 \frac{(\alpha E_R)^{1/4}t^{-3/4}}
 {|(z_s-z_R)(z_s-z_R^*)|}
 \exp\left(-\sqrt{2\alpha E_Rt}\right).
 \label{eq:supp-essential-background}
\end{equation}
The corresponding pole contribution has magnitude
\begin{equation}
 |P_\alpha(t)|
 =
 \frac{\pi}{\gamma\mathcal N_\alpha}
 \exp\left[
 -\alpha E_R\operatorname{Re}\frac{1}{z_R}
 -\gamma t
 \right].
 \label{eq:supp-essential-pole-magnitude}
\end{equation}

For fixed $\alpha$, the pole eventually decays faster than the threshold background. To obtain increasing pole dominance, we increase $\alpha$ and move the fitting window to later times according to $t=c\alpha E_R$. The saddle point then becomes $z_s(c)=e^{-i\pi/4}/\sqrt{c}$, independent of $\alpha$, while the power-law factor in Eq.~\eqref{eq:supp-essential-background} becomes $(\alpha E_R)^{1/4}t^{-3/4}=c^{-3/4}(\alpha E_R)^{-1/2}$. Dividing Eq.~\eqref{eq:supp-essential-background} by Eq.~\eqref{eq:supp-essential-pole-magnitude} therefore gives
\begin{equation}
 \left|\frac{B_\alpha}{P_\alpha}\right|
 \sim
 \frac{\gamma c^{-3/4}}
 {\sqrt{\pi}\,
 |[z_s(c)-z_R][z_s(c)-z_R^*]|}
 (\alpha E_R)^{-1/2}e^{-\alpha\mu(c)},
 \qquad
 \mu(c)=E_R\left(
 \sqrt{2c}-\operatorname{Re}\frac{1}{z_R}-\gamma c
 \right).
 \label{eq:supp-essential-scaled}
\end{equation}

For a scaled fitting window in which $\mu(c)>0$, Eq.~\eqref{eq:supp-essential-scaled} shows that $B_\alpha/P_\alpha$ converges exponentially to zero as $\alpha$ increases. The bias relation in Eq.~\eqref{eq:supp-fit-bias} then implies that the fitted pole $E_{\rm fit}$ converges exponentially to $z_R$.

We test Eq.~\eqref{eq:supp-essential-scaled} using the continuous-space Winter model \cite{Winter:1961},
\begin{equation}
 H=-\frac{d^2}{dr^2}+g\delta(r-R),\qquad r>0,
\end{equation}
in units where $\hbar^2/(2m)=1$, with $R=8$ and $gR=30$. We introduce the dimensionless energy and time
$\epsilon=ER^2$ and $s=t/R^2$, and denote the dimensionless pole by
$\zeta_R=z_RR^2=e_R-i\gamma_R$. For the lowest resonance,
$\zeta_R=9.25063908388483-0.059372088073924\,i$.

For each $\alpha>0$, we use the spectral weighting function
$F_\alpha(\epsilon)=\exp(-\alpha e_R/\epsilon)$, calculate the corresponding survival amplitude, and fit it to
$Z_{\rm fit}e^{-i\epsilon_{\rm fit}s}$ using
Eq.~\eqref{eq:supp-log-fit}. The scaled fitting window is
$c=s/(\alpha e_R)\in[0.9c_\star,1.1c_\star]$, where
$c_\star=1/(2\gamma_R^2)$. Equation~\eqref{eq:supp-essential-scaled} then gives
\begin{equation}
 \mu_{\rm pred}
 =\min_{c\in[0.9c_\star,1.1c_\star]}\mu(c)
 =76.6988279,
 \qquad
 \frac{|\Delta E|}{|z_R|}
 \equiv
 \frac{|E_{\rm fit}-z_R|}{|z_R|}
 =
 \frac{|\epsilon_{\rm fit}-\zeta_R|}{|\zeta_R|}
 \sim{\rm poly}(\alpha)e^{-\mu_{\rm pred}\alpha}.
 \label{eq:supp-winter-error}
\end{equation}
The local-energy variance is evaluated on the same time grid.

\begin{figure}[t]
 \centering
 \includegraphics[width=0.73\linewidth]
 {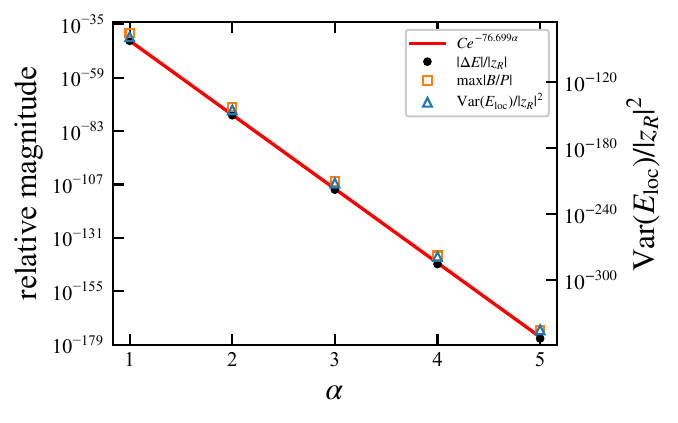}
 \caption{\label{fig:supp-essential-precision}
 Relative error of the resonance pole extraction for the Winter model.
 The fitted-pole errors track the
 exact contour values of $\max|B/P|$ over more than 170 decimal digits.
 The red curve uses the steepest-descent prediction in
 Eq.~\eqref{eq:supp-essential-scaled}, normalized at the first displayed
 point, $\alpha=1$. Filled black circles show the pole error, open orange
 squares show $\max|B/P|$, and open blue triangles show
 $\operatorname{Var}(E_{\rm loc})/| z_R|^2$ on the right axis. All three
 diagnostics use the same fitting grids.}
\end{figure}

As shown in Fig.~\ref{fig:supp-essential-precision}, the fitted-pole error and
$\max|B/P|$ follow the predicted exponential dependence on $\alpha$. A fit
for $\alpha\geq2$ gives $\mu_{\rm num}=77.0764$, differing from
$\mu_{\rm pred}$ by $0.49\%$, while the relative pole error reaches
$5.1\times10^{-177}$ at $\alpha=5$. The simultaneous decrease of the
local-energy variance confirms that suppressing the background produces an
increasingly constant local energy over the scaled fitting window.

\subsubsection{Generalization to the lattice}
\label{subsec:supp-finite-lattice}

We now extend the spectral-overlap construction analysis to a finite-volume lattice Hamiltonian. Poisson summation separates the survival amplitude into an infinite-volume pole contribution, a spectral background, and finite-volume images. We then use the steepest descent method to determine their large-$L$ behavior and verify the predicted convergence pattern using numerical calculation.

For the one-dimensional lattice benchmark, the scattering band
$E(k)=1-\cos k$, with $0<k<\pi$, extends from the threshold $E=0$ to the upper band edge $E_{\max}=2$. To suppress both endpoints, we define
\begin{equation}
 x(E)=
 \frac{E/(E_{\max}-E)}
 {E_R/(E_{\max}-E_R)},
 \qquad
 F_{\alpha,\beta}(E)
 =\exp\left[-\frac{\alpha}{x(E)}-\beta x(E)\right],
 \qquad \alpha,\beta>0.
 \label{eq:supp-band-coordinate}
\end{equation}
Here $x(E_R)=1$, while $x\to0$ at the threshold and $x\to\infty$ at the upper band edge. For $E_{\max}=2$,
$x(k)=\tan^2(k/2)/r$, where $r=E_R/(2-E_R)$.


Open boundary conditions quantize the band momenta through
$Q_L(k_n)=n\pi$, where $Q_L(k)=(L+1)k+\delta(k)$ and $\delta(k)$ is the
scattering phase shift of the interior potential.  The level density is
$dn/dk=Q_L'(k)/\pi$, so the energy cell of level $n$ is
\begin{equation}
 \Delta E_n=
 \frac{dE/dk}{dn/dk}\bigg|_{k_n}
 =\frac{\pi\sin k_n}{L+1+\delta'(k_n)}.
 \label{eq:supp-lattice-quantization}
\end{equation}
For the band eigenpairs $\{E_n,|E_n\rangle\}$ with $0<E_n<2$, we define
\begin{equation}
 p_n =\frac{1}{\mathcal N_L}
 \frac{\Delta E_n F_{\alpha,\beta}(E_n)}
 {(E_n-z_R)(E_n-z_R^*)}, \qquad 
 |\Psi_{\alpha,\beta,L}\rangle =\sum_n\sqrt{p_n}|E_n\rangle,
 \qquad A_L(t)=\sum_np_ne^{-iE_nt}.
 \label{eq:supp-lattice-state}
\end{equation}
Because $p_n>0$ and $\sum_np_n=1$ (the constant $\mathcal N_L$ is chosen to ensure the normalization), this is a normalized state that evolves under
the original finite Hermitian Hamiltonian. It is the lattice analog of
Eq.~\eqref{eq:supp-spectral-state}.

To separate the infinite-volume and finite-volume contributions, we apply Poisson summation to the quantization condition,
\begin{equation}
 \sum_n\delta(k-k_n)
 =\frac{Q_L'(k)}{\pi}
 \sum_{m\in\mathbb Z}e^{2imQ_L(k)},
 \label{eq:supp-lattice-comb}
\end{equation}
where $Q_L'(k)>0$ for the band levels considered here. The factor $Q_L'(k)/\pi$ in Eq.~\eqref{eq:supp-lattice-comb} cancels the factor $\pi/Q_L'(k_n)$ in $\Delta E_n$, leaving $dE/dk=\sin k$. The normalized survival amplitude can therefore be written exactly as
\begin{align}
 A_L(t)
 &=\frac{1}{\mathcal N_L}
 \sum_n\Delta E_n
 \frac{F_{\alpha,\beta}(E_n)e^{-iE_nt}}
 {(E_n-z_R)(E_n-z_R^*)}
 =\frac{1}{\mathcal N_L}
 \sum_{m\in\mathbb Z}I_m(t),
 \label{eq:supp-lattice-poisson}\\
 I_m(t)
 &=\int_0^\pi dk\,
 \frac{\sin k\,F_{\alpha,\beta}(E(k))e^{-iE(k)t}}
 {(E(k)-z_R)(E(k)-z_R^*)}
 e^{2imQ_L(k)},
 \label{eq:supp-lattice-poisson-image}
\end{align}
where $\mathcal N_L=\sum_{m\in\mathbb Z}I_m(0)$ ensures $A_L(0)=1$. The zero Poisson mode $I_0$ is the corresponding infinite-volume band integral and is independent of $L$ and $\delta(k)$, while the nonzero modes contain the finite-volume quantization effects. Since $e^{2imQ_L(k)}=e^{2im(L+1)k}S(k)^m$, with $S(k)=e^{2i\delta(k)}$, the nonzero modes can be interpreted as finite-volume return contributions, with $|m|$ labeling the return order.

For $m=0$, the $I_0(t)$ in Eq.~\eqref{eq:supp-lattice-poisson-image} reduces to the corresponding infinite-volume band integral. Its normalized survival amplitude is therefore $A_\infty(t)=I_0(t)/I_0(0)$. We define the remaining finite-volume contribution by
\begin{equation}
 \delta A_{\rm FV}(t)
 \equiv A_L(t)-A_\infty(t).
 \label{eq:supp-lattice-fv-exact}
\end{equation}

To separate the zero Poisson mode into pole and background contributions, we deform the integration interval $[0,\pi]$ into a contour $\Gamma_B$ with the same endpoints. The contour passes below $k_R=\arccos(1-z_R)$, where the branch is chosen such that $\operatorname{Im}k_R<0$, and crosses no other singularity. Since $E'(k_R)=\sin k_R$ and $z_R-z_R^*=-2i\gamma$, the residue theorem gives
\begin{equation}
\begin{aligned}
I_0(t)&=P_0(t)+B_0(t),\\
P_0(t)&=\frac{\pi}{\gamma}
F_{\alpha,\beta}(z_R)e^{-iz_Rt},\\
B_0(t)&=\int_{\Gamma_B}dk\,
\frac{\sin k\,F_{\alpha,\beta}(E(k))e^{-iE(k)t}}
{(E(k)-z_R)(E(k)-z_R^*)}.
\end{aligned}
\label{eq:supp-lattice-pole-term}
\end{equation}
Dividing by $I_0(0)$ defines $P(t)=P_0(t)/I_0(0)$ and
$B(t)=B_0(t)/I_0(0)$. Consequently,
\begin{equation}
A_L(t)=P(t)+B(t)+\delta A_{\rm FV}(t).
\label{eq:supp-lattice-amplitude-decomposition}
\end{equation}
This is the finite-volume counterpart of
Eq.~\eqref{eq:supp-essential-contour}, with the additional correction
$\delta A_{\rm FV}(t)$.

The large-$L$ behavior of the background and finite-volume contributions can be obtained using the same steepest-descent method as in Eq.~\eqref{eq:supp-steepest-rule}. We jointly scale the spectral weighting function and the fitting time according to
\begin{equation}
\alpha=\sigma_-L,\qquad
\beta=\sigma_+L,\qquad
t=\tau L+u,\qquad
u=O(1),
\label{eq:supp-lattice-trajectory-general}
\end{equation}
and write $Q_L(k)=Lk+\chi(k)$, where $\chi(k)=k+\delta(k)=O(1)$. Each Poisson mode then takes the form
\begin{equation}
I_m(\tau L+u)=\int_0^\pi G_m(k)e^{L\Phi_m(k)}e^{-iE(k)u}\,dk,
\label{eq:supp-lattice-action}
\end{equation}
where
\begin{equation}
G_m(k)=
\frac{\sin k\,e^{2im\chi(k)}}
{(E(k)-z_R)(E(k)-z_R^*)},
\qquad
\Phi_m(k)=
-\frac{\sigma_-}{x(k)}-\sigma_+x(k)
-i\tau E(k)+2imk.
\end{equation}

The saddle points of $\Phi_m(k)$ satisfy
\begin{equation}
x'(k_{m,s})\left[\frac{\sigma_-}{x(k_{m,s})^2}-\sigma_+\right]-i\tau\sin k_{m,s}+2im=0,\qquad
x'(k)=\frac{\tan(k/2)\sec^2(k/2)}{r}.
\label{eq:supp-lattice-saddle-equation}
\end{equation}
The contributing solutions are selected by the contour deformation. For $m\neq0$, the Poisson factor satisfies $|e^{2imLk}|=e^{-2mL\operatorname{Im}k}$ and decreases in the upper half-plane for $m>0$ and the lower half-plane for $m<0$. At a nondegenerate saddle point, where $\Phi_m''(k_{m,s})\neq0$, expanding $\Phi_m(k)$ to second order and applying Eq.~\eqref{eq:supp-steepest-rule} gives
\begin{equation}
\left|\frac{I_{m,s}}{P_0}\right|
\sim C_{m,s}(u)L^{-1/2}e^{-\mu_{m,s}L},\qquad
\mu_{m,s}=\operatorname{Re}\Phi_0(k_R)-\operatorname{Re}\Phi_m(k_{m,s}).
\label{eq:supp-action-gap}
\end{equation}
Here $I_{m,s}$ denotes the contribution from the saddle $k_{m,s}$. The $m=0$ saddles determine the background-to-pole ratio, while the $m\neq0$ saddles determine the finite-volume return contributions relative to the pole. For a fixed-width fitting window centered at $t=\tau L$, Eq.~\eqref{eq:supp-fit-bias} shows that the fitted-pole error has the same $L^{-1/2}e^{-\mu_{m,s}L}$ dependence. We define $\mu_{\rm bg}=\min_s\mu_{0,s}$ and $\mu_{\rm FV}=\min_{m\neq0,s}\mu_{m,s}$ over the contributing saddles. The smaller positive exponent controls the leading error at large $L$.

\begin{figure}[t]
 \centering
 \includegraphics[width=0.73\linewidth]
 {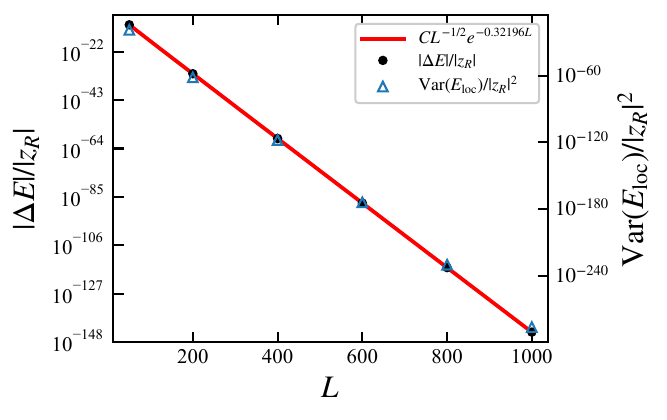}
 \caption{\label{fig:supp-lattice-scaling}
 Jointly scaled finite-lattice trajectory with the fixed-width fit in
 Eq.~\eqref{eq:supp-lattice-prescription}.  The red curve is
 $CL^{-1/2}e^{-\mu_{\rm up}L}$, normalized at $L=50$.  Filled black circles show
 $|E_{\rm fit}-z_R|/|z_R|$, and open blue triangles show
 $\operatorname{Var}(E_{\rm loc})/|z_R|^2$ on the same time grid.}
\end{figure}

We test these predictions using the short-range 1D two-body delta-site model defined in Eq.~\eqref{eq:supp-model-1d2bdelta}. For this model, the relevant contour deformations reach the contributing saddles without crossing a singularity of $G_m$. Figure~\ref{fig:supp-lattice-scaling} shows the relative fitted-pole error using
\begin{equation}
t_c=L,\qquad
\Delta t=13.4,\qquad
\alpha=\beta=0.12L.
\label{eq:supp-lattice-prescription}
\end{equation}
The two contributing $m=0$ saddle points, associated with the threshold and the upper band edge, are $k_{\rm low}=0.9651941-0.4169967i$ and $k_{\rm up}=2.2155162-0.3928686i$, with $\mu_{\rm low}=0.3579095$ and $\mu_{\rm up}=0.3219586$. The leading finite-volume saddle lies in the $m=1$ sector at $k_{1,s}=2.4179966+0.5187523i$ and gives $\mu_{\rm FV}=0.5362913$. The remaining nonzero Poisson modes have larger suppression exponents. The hierarchy $\mu_{\rm up}<\mu_{\rm low}<\mu_{\rm FV}$ therefore predicts the background-dominated behavior $CL^{-1/2}e^{-\mu_{\rm up}L}$. A fit to the full discrete calculation for $L\geq200$ gives $\mu_{\rm num}=0.3226732$, differing from $\mu_{\rm up}$ by $0.22\%$. The relative pole error and $\sqrt{\operatorname{Var}(E_{\rm loc})}/|z_R|$ decrease with the same exponential rate. The fitted pole therefore converges exponentially to $z_R$ as $L$ increases.

\begin{figure}[t]
\centering
\includegraphics[width=0.92\linewidth]
{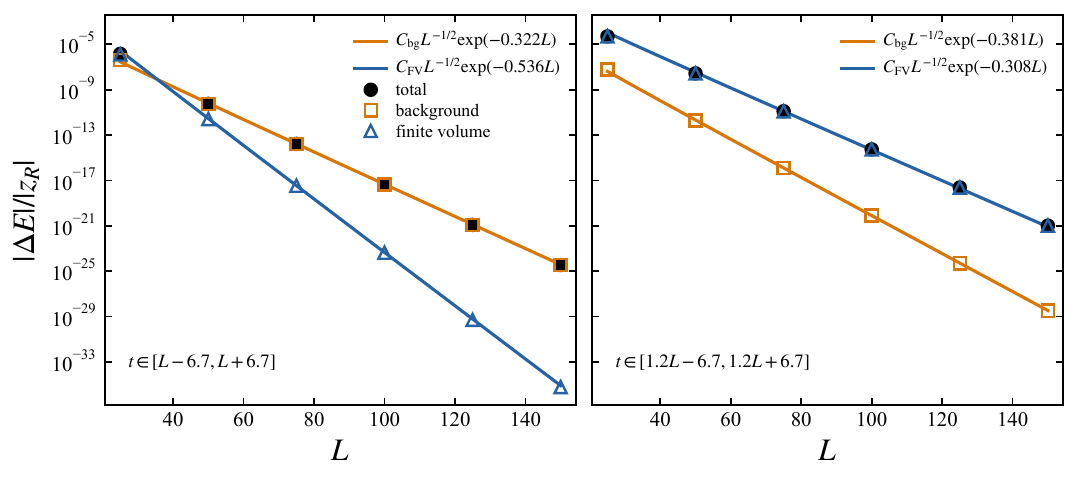}
\caption{\label{fig:supp-lattice-wall-background}
Separation of the fitted-pole error into spectral background and finite-volume return contributions for the two fixed-width fitting windows indicated in the panels. Filled black circles show $|E_{\rm fit}[A_L]-z_R|/|z_R|$. Open orange squares and open blue triangles show $|\delta E_{\rm bg}|/|z_R|$ and $|\delta E_{\rm FV}|/|z_R|$, respectively, as defined in Eq.~\eqref{eq:supp-lattice-fit-decomposition}. The solid lines show $L^{-1/2}e^{-\mu_{\rm bg}L}$ and $L^{-1/2}e^{-\mu_{\rm FV}L}$ using the independently calculated suppression exponents. Each line is normalized to its corresponding data point at $L=75$.}
\end{figure}

To determine whether the spectral background or the finite-volume correction in Eq.~\eqref{eq:supp-lattice-amplitude-decomposition} controls the fitted-pole error, Fig.~\ref{fig:supp-lattice-wall-background} separates their contributions to the complex fitted-pole error. For each $L=25,50,\ldots,150$, we evaluate $A_\infty(t)$ independently using 4096-point midpoint quadrature of Eq.~\eqref{eq:supp-lattice-poisson-image} with $m=0$. Doubling the number of quadrature points changes the fitted pole obtained from $A_\infty(t)$ by less than $6\times10^{-47}$ at $L=150$. Applying the 17-point complex-log fit in Eq.~\eqref{eq:supp-log-fit} to $A_L(t)$ and $A_\infty(t)$ gives the exact decomposition
\begin{equation}
\begin{aligned}
E_{\rm fit}[A_L]-z_R
&=\delta E_{\rm bg}+\delta E_{\rm FV},\\
\delta E_{\rm bg}
&\equiv E_{\rm fit}[A_\infty]-z_R,\\
\delta E_{\rm FV}
&\equiv E_{\rm fit}[A_L]-E_{\rm fit}[A_\infty].
\end{aligned}
\label{eq:supp-lattice-fit-decomposition}
\end{equation}
Both panels use $\alpha=\beta=0.12L$ and the fixed width $\Delta t=13.4$. For the window centered at $t_c=L$, the suppression exponents are $(\mu_{\rm low},\mu_{\rm up},\mu_{\rm FV})=(0.3579095,0.3219586,0.5362913)$, so the upper-band background dominates. For the window centered at $t_c=1.2L$, the two background suppression exponents are $0.4227615$ and $0.3809889$, while the leading $m=1$ saddle at $k_{1,s}=1.5726223+0.8413903i$ gives $\mu_{\rm FV}=0.3078398$. The finite-volume return contribution therefore dominates.

\subsection{Benchmark models and reference values for the resonance poles}
\label{sec:supp-benchmarks}

This section specifies the four lattice Hamiltonians used in the main text and explains how their reference resonance poles are determined. The lattice Schr\"odinger equation is used for the two 1D two-body models, while CAP calculations provide the references for the 3D two-body and 1D three-body models.


The four lattice models considered in the main text are defined as follows:
\begin{itemize}
\item
{\it Two particles interacting via a short-range potential in 1D}\\
This model is used for the precision calculations visualized in
Fig.~\ref{fig:psm-mp-comparison}. We work in the normalized even-parity
basis $|0\rangle_+=|0\rangle$ and
$|n\rangle_+=(|n\rangle+|-n\rangle)/\sqrt{2}$, where $n=1,\ldots,L$ is
the lattice-site distance from the origin. In this basis,
\begin{equation}
 V_n=-4\delta_{n0}+3\delta_{n2},\qquad a=1,\qquad \mu=1.
 \label{eq:supp-model-1d2bdelta}
\end{equation}
The complete lattice Hamiltonian is given in
Eq.~\eqref{eq:supp-lse-delta-hamiltonian}.
\item
{\it Two particles interacting via long- and short-range potentials in 1D}\\
We consider two particles on a half-line ($n=1,...,L/a$) interacting via the potential 
\begin{equation}
 V(r)=-8e^{-0.16r^2}+4e^{-0.04r^2}+\frac{1}{r},
 \qquad r=na,\qquad a=0.25,\qquad \mu=1,
 \label{eq:supp-model-1d2b}
\end{equation}
which features both the short-range Gaussian and the long-range Coulomb-like interactions. The results of our calculations using the PSM are shown in Fig.~\ref{fig:surv_amps}(a). 
\item
{\it Two particles interacting via long- and short-range potentials in 3D}\\
Here, we use the central interaction from the previous model on a
3D cubic lattice with spacing $a=1$. At the origin, the Coulomb term is
regularized as $1/(r+a/2)$. We then reduce the Hamiltonian to the $A_1$
radial-shell basis as described in Sec.~\ref{sec:supp-a1}. Our results for this model are presented in Fig.~\ref{fig:surv_amps}(b). 
\item
{\it Three particles in 1D}\\
We consider three particles of mass
$m=938.92\,\mathrm{MeV}$ moving in one dimension. After removing the center of
mass, we use Jacobi coordinates $y_1=x_2-x_1$ and $y_2=x_3-(x_1+x_2)/2$, where $x_i$ are the particle positions, with lattice spacing
$1\,\mathrm{fm}$. The interaction contains repulsive pairwise Gaussians and
an attractive three-body Gaussian,
\begin{align}
 V_{\mathrm{2B}}={}&V_+\left[e^{-y_1^2/R_+^2}
 +e^{-(y_2-y_1/2)^2/R_+^2}+e^{-(y_2+y_1/2)^2/R_+^2}\right],\nonumber\\
 V_{\mathrm{3B}}={}&-V_-\exp\!\left[-\frac{y_1^2+(y_2-y_1/2)^2
 +(y_2+y_1/2)^2}{R_-^2}\right],
 \label{eq:supp-model-1d3b}
\end{align}
with
$V_+=30\,\mathrm{MeV}$, $R_+=1\,\mathrm{fm}$,
$V_-=100\,\mathrm{MeV}$, $R_-=2\,\mathrm{fm}$, and $\hbar c=197.327\,\mathrm{MeV\,fm}$.
The reduced masses are $\mu_1=m/2$ and $\mu_2=2m/3$.
Our results for this model are shown in Fig.~\ref{fig:surv_amps}(c). 
\end{itemize}

\subsubsection{Resonance poles from the lattice Schr\"odinger equation}

On the half-line lattice \(r_n=na\), with
\(t_{\mathrm{hop}}=(2\mu a^2)^{-1}\), the interior equation is
\begin{equation}
 -t_{\mathrm{hop}}u_{n-1}
 +\left(2t_{\mathrm{hop}}+V_n-E\right)u_n
 -t_{\mathrm{hop}}u_{n+1}=0.
 \label{eq:supp-lse-lattice}
\end{equation}
At sites $n$ where \(u_n\neq0\), the ratio \(q_n=u_{n+1}/u_n\) obeys
\begin{equation}
 q_n=\frac{2t_{\mathrm{hop}}+V_n-E}{t_{\mathrm{hop}}}
 -\frac{1}{q_{n-1}},
 \qquad
 q_{n-1}=
 \frac{1}{(2t_{\mathrm{hop}}+V_n-E)/t_{\mathrm{hop}}-q_n}.
 \label{eq:supp-lse-riccati}
\end{equation}
The boundary condition fixes a regular solution up to normalization:
\(u_0=0\) for the radial problem. Its ratio \(q_n^{\mathrm{reg}}\) is propagated
outward to a matching site \(M\). An outgoing solution determines a second
ratio \(q_n^{\mathrm{out}}\), propagated inward to the same site $M$.

At sites where both the regular and outgoing solutions obey Eq.~\eqref{eq:supp-lse-lattice}, their discrete
Wronskian
\begin{equation}
 W_n=u_n^{\mathrm{reg}}u_{n+1}^{\mathrm{out}}
     -u_{n+1}^{\mathrm{reg}}u_n^{\mathrm{out}}
\end{equation}
satisfies \(W_n=W_{n-1}\). Provided
\(u_M^{\mathrm{reg}}u_M^{\mathrm{out}}\neq0\),
\[
 W_M=-u_M^{\mathrm{reg}}u_M^{\mathrm{out}}F(E),
 \qquad
 F(E)=q_M^{\mathrm{reg}}(E)-q_M^{\mathrm{out}}(E),
\]
so the pole condition is
\begin{equation}
 F(E)=0.
\end{equation}
At such an energy, the regular and outgoing solutions are linearly
dependent, so one nonzero lattice wave function satisfies both boundary
conditions.


For a strictly finite-range potential, \(V_n=0\) for \(n>M\), the outgoing
solution is \(u_n^{\mathrm{out}}\propto\lambda^n\), with
\(\lambda=e^{ik_{\mathrm{lat}}}\). Substitution into the free lattice equation
gives
\begin{equation}
 E=t_{\mathrm{hop}}\left(2-\lambda-\lambda^{-1}\right),
 \qquad q_M^{\mathrm{out}}=\lambda.
 \label{eq:supp-lse-short-range}
\end{equation}
A decaying resonance belongs to the branch
\(\operatorname{Im}k_{\mathrm{lat}}<0\) with \(|\lambda|>1\), so the
matching condition \(F(E)=0\) becomes
\(q_M^{\mathrm{reg}}(E)=\lambda(E)\).

For the delta-site well-and-barrier benchmark of
Eq.~\eqref{eq:supp-model-1d2bdelta},
\begin{equation}
 t_{\mathrm{hop}}=\frac12,\qquad
 H_{nn}=1+V_n,\qquad H_{01}=-\frac1{\sqrt2},\qquad
 H_{n,n+1}=-\frac12\ (n\geq1),\qquad
 V_n=-4\delta_{n0}+3\delta_{n2}.
 \label{eq:supp-lse-delta-hamiltonian}
\end{equation}
The even-parity boundary equation gives
\(q_0=\sqrt2(1+V_0-E)\), while the equations at \(n=1,2\) give
\(q_1=2(1+V_1-E)-\sqrt2/q_0\) and
\(q_2=2(1+V_2-E)-1/q_1\). Defining
\(S=\lambda+\lambda^{-1}\), so that \(E=1-S/2\), and inserting
\(V_0=-4\), \(V_1=0\), and \(V_2=3\), these ratios become
\(q_0=(S-8)/\sqrt2\),
\(q_1=S-2/(S-8)\), and
\(q_2=6+S-(S-8)/(S^2-8S-2)\).
Since \(V_n=0\) for \(n>2\), choosing \(M=2\) makes the outgoing condition
\(q_2=\lambda\). Substituting \(S=\lambda+\lambda^{-1}\) and clearing the
denominators gives
\begin{equation}
 6\lambda^5-48\lambda^4-49\lambda^2-2\lambda+1=0.
 \label{eq:supp-lse-delta-model}
\end{equation}
The root on the decaying resonance branch gives
\begin{equation}
 E_{\mathrm{pole}}
 =
 1.042022724906348337200060388\dots
-0.013115400294386927209457867\dots\,\mathrm{i}.
 \label{eq:supp-lse-delta-pole}
\end{equation}

For the Gaussian-plus-Coulomb benchmark, the exponentially small Gaussian
terms are neglected for \(n>M\), while the nonvanishing Coulomb tail
\(V_n=Z/(na)\) is retained. Here \(M\) is the site where the regular and
outgoing solutions are matched, \(N>M\) is the remote site where the outgoing
ratio is initialized, and \(p\) is the highest retained order in its
large-\(n\) expansion.

With \(E=2t_{\mathrm{hop}}(1-\cos k_{\mathrm{lat}})\),
\(\lambda=e^{ik_{\mathrm{lat}}}\),
\(A=2-E/t_{\mathrm{hop}}\), and \(c=Z/(at_{\mathrm{hop}})\), the ratio in the
Coulomb-only region obeys
\(q_n=A+c/n-1/q_{n-1}\). On the outgoing branch,
\(\operatorname{Im}k_{\mathrm{lat}}<0\) and \(|\lambda|>1\).

To construct the outgoing ratio at large \(n\), set \(x=1/n\) and write
\(q_n=Q(x)\sim\lambda+\sum_{j\geq1}b_jx^j\). Since
\(q_{n-1}=Q(1/(n-1))=Q(x/(1-x))\), the ratio equation becomes
\begin{equation}
 Q(x)=A+cx-\frac{1}{Q\!\left(x/(1-x)\right)}.
 \label{eq:supp-lse-discrete-boundary}
\end{equation}
The constant term selects the outgoing root \(\lambda\), and each higher
order determines \(b_j\) from the lower-order coefficients; for example,
\(b_1=c/(1-\lambda^{-2})\). The denominator vanishes at the lattice band
edges \(\lambda=\pm1\).

Defining
\(Q_p(x)=\lambda+\sum_{j=1}^{p}b_jx^j\), we initialize the outgoing ratio at
the remote site by \(q_N^{\mathrm{out}}\approx Q_p(1/N)\) and propagate it
inward through the Coulomb-only region. The regular ratio is propagated
outward with the full potential, and the pole is determined by
\begin{equation}
 F(E)=q_M^{\mathrm{reg}}(E)-q_M^{\mathrm{out}}(E)=0.
\end{equation}
Setting \(c=0\) makes all \(b_j=0\) and recovers the finite-range boundary
\(q_N^{\mathrm{out}}=\lambda\).

For \(M=240\), \(N=3200\), and \(p=28\), the
calculation is performed with 80-decimal arithmetic. The resulting lattice
reference is
\begin{equation}
 \begin{aligned}
 \operatorname{Re}E_{\mathrm{ref}}
 &=1.0817540526955831366272052150550567447200\dots,\\
 \operatorname{Im}E_{\mathrm{ref}}
 &=-0.0005362387759959912600385499130986323020\dots.
 \end{aligned}
 \label{eq:supp-lse-reference}
\end{equation}

Table~\ref{tab:lse_sweep_supp} varies one numerical parameter at a time,
while the other two are fixed at the reference values
\(M=240\), \(N=3200\), and \(p=28\). Each difference is measured relative
to Eq.~\eqref{eq:supp-lse-reference}. The pole is 
numerically stable to at least 40 decimal places throughout the tested
ranges.

\begin{table}[h]
\caption{\label{tab:lse_sweep_supp}
Convergence test for the resonance pole solution of the lattice
Schr\"odinger equation under variations of the remote boundary \(N\),
asymptotic order \(p\), and matching site \(M\).}
\begin{ruledtabular}
\begin{tabular}{ccc}
varied parameter & value & \(\lvert E-E_{\mathrm{ref}}\rvert\) \\
\colrule
\multicolumn{3}{c}{\(N\) variation with \(M=240\) and \(p=28\)}\\
\(N\) & 1600 & \(4.59\times10^{-63}\) \\
\(N\) & 2400 & \(4.16\times10^{-68}\) \\
\(N\) & 4000 & \(1.15\times10^{-71}\) \\
\(N\) & 4800 & \(1.15\times10^{-71}\) \\
\colrule
\multicolumn{3}{c}{\(p\) variation with \(M=240\) and \(N=3200\)}\\
\(p\) & 20 & \(9.03\times10^{-56}\) \\
\(p\) & 22 & \(7.45\times10^{-60}\) \\
\(p\) & 24 & \(7.34\times10^{-64}\) \\
\(p\) & 26 & \(8.51\times10^{-68}\) \\
\(p\) & 30 & \(1.15\times10^{-71}\) \\
\(p\) & 32 & \(1.15\times10^{-71}\) \\
\(p\) & 34 & \(1.15\times10^{-71}\) \\
\colrule
\multicolumn{3}{c}{\(M\) variation with \(N=3200\) and \(p=28\)}\\
\(M\) & 200 & \(1.38\times10^{-47}\) \\
\(M\) & 220 & \(9.27\times10^{-57}\) \\
\(M\) & 260 & \(8.45\times10^{-67}\) \\
\end{tabular}
\end{ruledtabular}
\end{table}

\subsubsection{Resonance poles using the CAP method}
\label{subsec:CAP}
The CAP benchmarks are obtained by diagonalizing
$H_{\rm CAP}=H-i\eta W$, with $\eta>0$ and $W\geq0$.
The absorption increases smoothly from zero at the origin
toward the coordinate boundaries. We construct $W$ from
the flat exponential profile 
\begin{equation}
 w(u)=
 \begin{cases}
 0, & u=0,\\
 \exp\!\left(1-u^{-2}\right), & 0<u\leq1,
 \end{cases}
 \qquad
 u_j=\frac{r_j}{L},
 \label{eq:supp-cap-profile}
\end{equation}

as
\begin{equation}
 W=\sum_j w(u_j),\qquad
 \{r_j\}=
 \begin{cases}
 \{r\}, & \text{1D two-body},\\
 \{|x|,|y|,|z|\}, & \text{3D two-body},\\
 \{|y_1|,|y_2|\}, & \text{1D three-body}.
 \end{cases}
 \label{eq:supp-cap-shape}
\end{equation}
In the 3D radial-shell calculation, $W$ is averaged over
the sites belonging to each shell as described in Sec.~\ref{sec:supp-a1}.

For each volume, the absorber strength $\eta$ is swept
over a logarithmic grid. We follow the eigenvalue trajectory
associated with the target resonance and select the point
of least sensitivity to $\eta$, measured numerically by
$|dE/d\ln\eta|$. The 1D Coulomb reference is instead
supplied by solving the lattice Schr\"odinger equation as described above. For the 3D two-body and 1D
three-body panels of Fig.~\ref{fig:surv_amps}, the large-volume CAP reference
poles are, respectively,
\begin{equation}
 \begin{aligned}
 z_R^{(3\mathrm{D},2\mathrm{B})}&=1.3826296-0.0057115\,i,\\
 z_R^{(1\mathrm{D},3\mathrm{B})}&=8.1846910-1.0135355\,i\ \mathrm{MeV}.
 \end{aligned}
 \label{eq:supp-cap-reference-poles}
\end{equation}

\subsubsection{Radial-shell projection for the 3D benchmark}
\label{sec:supp-a1}

For the 3D benchmark we use a reduced radial-shell basis on a cubic lattice.
Let $\mathbf n=(n_x,n_y,n_z)\in\mathbb Z^3$ label lattice sites and group sites
by the value
\begin{equation}
  R_i^2=n_x^2+n_y^2+n_z^2,\qquad r_i=a\sqrt{R_i^2}.
\end{equation}
Let ${\cal S}_i$ be the set of sites in shell $i$, with $N_i=|{\cal S}_i|$. We define the normalized shell state
\begin{equation}
  |i\rangle=\frac{1}{\sqrt{N_i}}
  \sum_{\mathbf n\in{\cal S}_i}|\mathbf n\rangle .
\end{equation}
Each shell state is invariant under the cubic group and lies in the $A_1$ sector. We keep one state for each distinct $R_i^2$, so this is a reduced radial subspace, not the full $A_1$ basis.

With $U_{\mathbf n i}=N_i^{-1/2}$ on shell $i$, the reduced Hamiltonian is
$H_{\rm shell}=U^\dagger H U$. If $B_{ij}$ is the number of nearest-neighbor
bonds between shells $i$ and $j$, then
\begin{equation}
  (H_{\rm shell})_{ii}=6t_{\rm hop}+V(r_i),\qquad
  (H_{\rm shell})_{ij}
  =-t_{\rm hop}\frac{B_{ij}}{\sqrt{N_iN_j}}\quad(i\ne j),
\end{equation}
with $t_{\rm hop}=1/(2\mu a^2)$. These matrix elements follow directly from
the normalized shell states and are the ones used in the 3D benchmark.

\subsection{Compact-state finite-size effects}
\label{sec:supp-psm-scaling}

The spectral construction discussed in Sec.~\ref{subsec:SpectralConstruction} proves exponential convergence for a constructed family of normalizable states. However, the PSM uses strictly compact states, so this proof does not apply directly. Below, we prove a simpler result for the 1D nearest-neighbor example: free propagation through the exterior is exponentially small when no real lattice momentum can cover the required distance in the available time.

\subsubsection{One-dimensional nearest-neighbor example}
Consider first an infinite 1D nearest-neighbor lattice with spacing $a$ and mass $m$,
\begin{equation}
 E(k)=\frac{1-\cos(ka)}{ma^2},\qquad
 v_g(k)=\frac{\sin(ka)}{ma},\qquad
 v_{\max}=\frac{1}{ma}
 \label{eq:supp-psm-vmax}
\end{equation}
where $E(k)$ gives the dispersion relation, $v_g$ the group velocity, and $v_{\rm max}$ the maximum group velocity. 
The free propagator over $n=D/a$ lattice spacings is, apart from an overall phase,
\begin{equation}
 K_n(t)=i^nJ_n(\widetilde t),\qquad
 \widetilde t=\frac{t}{ma^2}.
 \label{eq:supp-psm-bessel}
\end{equation}
Its magnitude can be bounded directly from the momentum integral. Shifting the contour by $i\eta$, with $\eta>0$, one gets
\begin{equation}
 |K_n(t)|\leq \exp[-n\eta+\widetilde t\sinh\eta].
 \label{eq:supp-psm-contour-bound}
\end{equation}
We now define
\begin{equation}
 q=\frac{D/t}{v_{\max}}=\frac{maD}{t}.
 \label{eq:supp-psm-q}
\end{equation}
When $q>1$, no real momentum has enough group velocity to cover $D$ in time $t$. The optimal contour shift satisfies $\cosh\eta=q$, which gives
\begin{equation}
 |K_n(t)|\leq
 \exp\left[-\frac{D}{a}\operatorname{arcosh}q
 +\frac{t}{ma^2}\sqrt{q^2-1}\right].
 \label{eq:supp-psm-return-bound}
\end{equation}
The amplitude outside the group-velocity cone is not exactly zero. It comes from complex momentum and is exponentially small.

We now scale distance and time with a dimensionless linear size $L$,
\begin{equation}
 D=\delta La,\qquad t=\tau ma^2L .
\end{equation}
Then, $q=\delta/\tau$ is fixed, and
\begin{equation}
 |K_n(t)|\leq e^{-c_{\rm ret}L},\qquad
 c_{\rm ret}=\delta\operatorname{arcosh}q-\tau\sqrt{q^2-1}>0 .
 \label{eq:supp-psm-linear-bound}
\end{equation}
The positivity follows from
$c_{\rm ret}=\tau[q\operatorname{arcosh}q-\sqrt{q^2-1}]$ for $q>1$.
Thus, if the observation window ends a fixed fraction before the fastest real-momentum round trip, the free exterior propagation needed for a return is exponentially small in the linear system size. Local interactions in the compact region can change amplitudes and prefactors, but any path to a distant wall must still include this exterior propagation. As $q\to1^+$, the complex momentum approaches the real axis and $c_{\rm ret}\to0$. This derivation is valid for the 1D nearest-neighbor example; other lattices require their own dispersion relations.

For the 1D benchmark in Fig.~\ref{fig:psm-mp-comparison},
we have $a=ma^2=1$, $L_{\rm init}=0.2L$, and the latest fitted time is
$t=1.1L$.  A returning path emitted from a site
$n\leq L_{\rm init}$ travels to the wall and back, so that
\begin{equation}
 D=2(L-n)\in[1.6L,2L].
 \label{eq:supp-psm-distance-range}
\end{equation}
Because the optimized coefficients are extremely small near the
edge of the support, the effective return distance should be
larger than the lower endpoint.

Evaluating Eq.~\eqref{eq:supp-psm-linear-bound} at $\tau=1.1$ for the
two endpoints gives
\begin{equation}
 c_{\rm ret}\big|_{D=1.6L}=0.31108\ldots,
 \qquad
 c_{\rm ret}\big|_{D=2L}=0.73961\ldots .
 \label{eq:supp-psm-benchmark-return}
\end{equation}
An exponential fit to the PSM data gives
$c_{\rm PSM}=0.486$ in Fig.~\ref{fig:psm-mp-comparison}, which lies
between these two return-propagation rates. The observed convergence is,
therefore, consistent with exponentially suppressed finite-volume
returns.  A precise prediction would require both an
analytic form of the optimized PSM wave function and an evaluation of
its background-to-pole contribution; the state is presently obtained
numerically.

At fixed $m$, the continuum limit $a\to0$ sends
$v_{\max}=1/(ma)$ to infinity, so this lattice propagation bound
disappears, as expected for continuum Schr\"odinger evolution.  The
Winter-model result above is unaffected because it is an infinite-volume
spectral construction and does not rely on a maximum propagation
speed.

\subsubsection{Status of compact PSM convergence}
The resonance width fixes the outgoing flux and hence the matching
amplitude of an interior-normalized resonance, but it does not by itself
determine the finite-volume return exponent. For the nearest-neighbor
lattice above, let $k_R=\kappa-i\gamma_k$ and define $A$ as the outgoing
amplitude at a matching bond. The lattice current is
\begin{equation}
 j_{\rm out}
 =\frac{\sin(\kappa a)}{ma^2}|A|^2 .
\end{equation}
On the other hand,
\begin{equation}
 z_R=\frac{1-\cos(k_Ra)}{ma^2}
 =E_R-\frac{i\Gamma}{2},
\end{equation}
which gives
\begin{equation}
 \Gamma=
 \frac{2\sin(\kappa a)\sinh(\gamma_k a)}{ma^2}.
\end{equation}
With unit interior normalization, probability conservation requires
$j_{\rm out}=\Gamma$, and therefore
\begin{equation}
 |A|^2=2\sinh(\gamma_k a)
 =2\gamma_k a+\mathcal O\!\left[(\gamma_k a)^3\right].
 \label{eq:supp-psm-tail-amplitude}
\end{equation}
Thus, a narrow resonance state has a small outgoing matching amplitude, while
the exponential suppression of a finite-volume return is governed by
the lattice dispersion and propagation distance in
Eq.~\eqref{eq:supp-psm-linear-bound}. For the exactly compact optimized
states in Fig.~\ref{fig:psm-mp-comparison}, the full pole error is
numerically consistent with
\begin{equation}
 \frac{|E_{\rm PSM}-z_R|}{|z_R|}
 \sim {\rm poly}(L)e^{-c_{\rm PSM}L}.
 \label{eq:supp-psm-observed-exp}
\end{equation}
Here $L$ denotes the linear system size. A general proof of this law for
arbitrary strictly compact PSM-optimized states is not presently available.
The continuum spectral construction proves exponential convergence for a
normalizable family, while the finite-lattice spectral calculation gives a
saddle-point prediction consistent with the full discrete sum. Independently,
the one-dimensional nearest-neighbor bound proves exponential suppression of
the exterior propagation required for a boundary return.

\subsubsection{Higher-dimensional and many-body systems}
\label{sec:supp-universality}
In higher dimensions, $L$ should be interpreted as a characteristic
propagation or hyperradial length. The finite-size law $\exp(-cL)$ therefore
becomes $\exp(-cV^{1/d})$ for a $d$-dimensional volume $V$.

For an $N$-body system in $d$ spatial dimensions, removal of the center of mass leaves $N-1$ Jacobi vectors and an effective dimension $D=d(N-1)$. If a single democratic hyperradial channel dominates, the exterior resonance wave has the asymptotic form
\begin{equation}
 \Psi_R(\rho)\simeq
 A\,\rho^{-(D-1)/2}e^{i\kappa\rho}e^{\gamma\rho},
\end{equation}
where $K_R=\kappa-i\gamma$ is the complex hyperradial momentum. The power of $\rho$ changes algebraic factors but not the outgoing flux relation. With unit interior normalization and quadratic hyperradial dispersion, the flux through a matching hypersphere is proportional to $\kappa |A|^2$, while $\Gamma\propto\kappa\gamma$, so the squared matching amplitude is proportional to $\gamma$ for a single open channel.

This width dependence affects the outgoing-tail amplitude, not a universal exponential convergence length. In higher-dimensional or many-body problems, several thresholds, lattice band edges, sequential-decay channels, long-range couplings, and finite-volume paths may compete. For a given spectral construction, the smallest positive action gap among accessible saddles sets the leading rate. Propagation estimates require the appropriate multidimensional dispersion relation and path length. Multiple open channels or asymptotic momenta must be treated channel by channel.

\subsection{Finite-volume scaling of the CAP method}
\label{sec:supp-cap-scaling}

As described in Sec.~\ref{subsec:CAP}, the CAP method modifies the Hamiltonian near the finite-volume boundary. Its error is controlled by two reflected amplitudes: reflection generated by the onset of the absorber and reflection from the hard wall after incomplete absorption.

\subsubsection{Finite-volume effects from two reflections}
We take
\begin{equation}
    H_{\rm CAP}=H-i\eta W(u),
    \qquad
    W(u)=\exp\left(1-\frac{1}{u^2}\right),
    \qquad
    u=\frac{x}{L},
\end{equation}
for $0<x<L$, with $W(0)=0$ and $W(1)=1$. The resonance momentum is written as
$k_R=\kappa-i\gamma_k$.

We use $R$ to denote the magnitude of the reflected-to-incident wave amplitude measured in the free region before the absorber. For small reflection, the pole shift is linear in the reflected amplitude,
$
    |\Delta E_{\rm CAP}|
    \sim
    R_{\rm onset}+R_{\rm wall}.
$

\begin{enumerate}
    \item \textbf{Onset reflection}\\
    In the first Born approximation, back-reflection changes the momentum from $+k_R$ to $-k_R$. It is therefore controlled by the Fourier component of the CAP at momentum transfer $2k_R$,
    \begin{equation}
        R_{\rm onset}
        \propto
        \eta
        \left|
        \int dx\,W(x/L)e^{2ik_Rx}
        \right|.
    \end{equation}
    For the smooth activation region near $u=0$, the large-$L$ dependence is controlled by
    \begin{equation}
        J_{\rm onset}(L)
        \sim
        L\int du\,
        \exp\left[
            -\frac{1}{u^2}+2ik_RLu
        \right].
    \end{equation}
    Writing $S(u)=-u^{-2}+2ik_RLu$, the relevant saddle point satisfies
    \begin{equation}
        u_s=e^{i\pi/6}(k_RL)^{-1/3},
        \qquad
        S(u_s)
        =
        -3e^{-i\pi/3}(k_RL)^{2/3}.
    \end{equation}
    The Gaussian width around the saddle point scales as $L^{-2/3}$. Including the factor $L$ in $J_{\rm onset}$ therefore gives
    \begin{equation}
        R_{\rm onset}
        \sim
        A\,\eta L^{1/3}
        \exp\left(-cL^{2/3}\right),
        \qquad
        c=
        3\,{\rm Re}\!\left[
            e^{-i\pi/3}k_R^{2/3}
        \right],
        \label{eq:cap-onset}
    \end{equation}
    where $A$ is independent of $L$. The coefficient $c$ is fixed by the resonance pole $z_R$ through the lattice dispersion relation.

    \item \textbf{Wall reflection}\\
    Inside the absorber, the local complex momentum is
    \begin{equation}
        q(u)
        =
        \sqrt{k_R^2+2\mu i\eta W(u)}.
    \end{equation}
    In the WKB approximation, the outgoing wave behaves as
    $\psi\sim\exp[iL\int q(u)\,du]$, so its magnitude is controlled by ${\rm Im}\,q$.
    A wave reflected from the hard wall crosses the absorber twice, giving
    \begin{equation}
        R_{\rm wall}
        \sim
        C\exp\left[
            -2L\int_0^1
            {\rm Im}\,q(u)\,du
        \right],
    \end{equation}
    where $C$ is independent of $L$.

    For weak absorption and a narrow resonance,
    \begin{equation}
        q(u)
        \simeq
        k_R
        +
        i\frac{\mu\eta}{k_R}W(u)
        \simeq
        \kappa-i\gamma_k
        +
        i\frac{\mu\eta}{\kappa}W(u).
    \end{equation}
    Hence
    \begin{equation}
        {\rm Im}\,q(u)
        \simeq
        -\gamma_k
        +
        \frac{\mu\eta}{\kappa}W(u).
    \end{equation}
    The first term describes the spatial growth of the outgoing Gamow wave, while the second gives CAP attenuation. The returned amplitude therefore scales as
    \begin{equation}
        R_{\rm wall}
        \sim
        C\exp\left[(2\gamma_k-B\eta)L\right],
        \qquad
        B=
        \frac{2\mu}{\kappa}
        \int_0^1 W(u)\,du ,
        \label{eq:cap-wall}
    \end{equation}
    with $B$ independent of $L$.
\end{enumerate}

\subsubsection{Optimal absorber strength}
Equations~\eqref{eq:cap-onset} and \eqref{eq:cap-wall} give the positive error envelope
\begin{equation}
    \epsilon(L,\eta)
    =
    A\eta L^{1/3}e^{-cL^{2/3}}
    +
    C e^{(2\gamma_k-B\eta)L}.
\end{equation}
Minimizing with respect to $\eta$ gives
\begin{equation}
    \eta_{\rm opt}
    =
    \frac{2\gamma_k}{B}
    +
    \frac{c}{B}L^{-1/3}
    +
    \mathcal O\left(\frac{\ln L}{L}\right).
\end{equation}
Thus, the leading error is
\begin{equation}
    \min_{\eta}|\Delta E_{\rm CAP}|
    \sim
    \mathcal O\left(
        e^{-cL^{2/3}}
    \right).
    \label{eq:cap-final-scaling}
\end{equation}

\begin{figure}[t]
    \centering
     \includegraphics[width=0.62\linewidth]{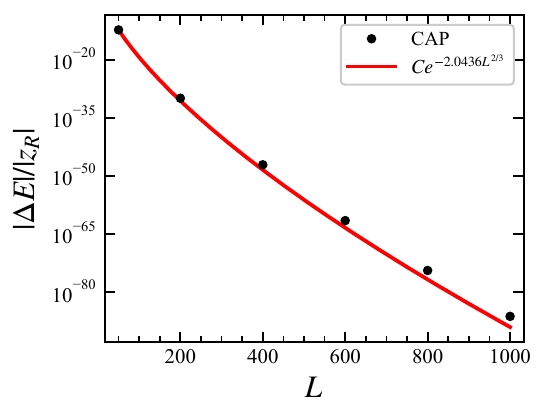}
    \caption{
    Relative error of the resonance pole using the finite-volume CAP method for the scaled flat absorber
    $W(u)=\exp(1-1/u^2)$.
    The solid line shows the theoretical leading scaling
    $C\exp[-cL^{2/3}]$, where
    $c=3\,{\rm Re}[e^{-i\pi/3}k_R^{2/3}]$
    is fixed by the resonance pole $z_R$ through the lattice dispersion relation.
    The single prefactor $C$ is fixed at the first numerical point.
    }
    \label{fig:cap-flat-error}
\end{figure}

Figure~\ref{fig:cap-flat-error} tests this leading stretched-exponential behavior for the model defined in Eq.~\eqref{eq:supp-lse-delta-hamiltonian}. The coefficient $c$ is obtained directly from Eq.~\eqref{eq:cap-onset} with resonance momentum $k_R$, while only the overall normalization is fixed from the numerical data.

 The stretched-exponential factor $2/3$ above is associated with the onset of the added absorber. On the other hand, complex scaling methods do not suffer from this drawback as they have no such onset term. Their only finite-volume error is generated by the remaining amplitude of the rotated resonance function at the truncation radius, which falls as $\exp[-2|k_R|L\sin(\theta-\theta_R)]$ for rotation angle $\theta$, with $\theta_R = \tfrac{1}{2}\arctan[\Gamma/(2E_R)]$. However, as already emphasized in the main text, complex scaling methods are not well suited to cases where the interaction is only available numerically.

\subsection{Broad resonances}
\label{sec:supp-psm-broad}

The benchmark calculation reported in Fig.~\ref{fig:psm-mp-comparison} corresponds to a narrow resonance with $\Gamma/E_R\approx0.025$. The PSM also converges exponentially for much broader resonances if the time-variance objective includes a mild guide term that keeps the search on the resonance branch. We use the same 1D short-range model as above (Sec.~\ref{sec:supp-benchmarks}), Eq.~\eqref{eq:supp-lse-delta-hamiltonian}, with barrier strength changed to $V_2=0.25$. The same finite-determinant method gives the certified pole
\begin{equation}
 E_{\rm pole}=0.76050892459385084343797289-0.36877043754703536725709550\,i.
 \label{eq:supp-psm-broad-pole}
\end{equation}

With the plain time-variance objective, the optimizer may instead find a quasi-eigenstate with $\mathrm{Im}\,E_{\rm loc}\approx0$. For a narrow resonance this state is well separated in energy and does not compete, but for a broad resonance it can lie close enough to do so. We therefore add a term that keeps the fitted $(\mathrm{Re}\,\bar E,\mathrm{Im}\,\bar E)$ within a broad tolerance region around a coarse, pole-blind estimate $(\mathrm{Re}\,\hat E,\mathrm{Im}\,\hat E)$,
\begin{equation}
 \mathcal L=\sqrt{\mathrm{Var}_t[E_{\rm loc}]+w\left[g(\Delta_{\rm Re})^2+g(\Delta_{\rm Im})^2\right]+\varepsilon^2}-\varepsilon,
 \label{eq:supp-psm-guide}
\end{equation}
with $\Delta_{\rm Re}=\mathrm{Re}\,\bar E-\mathrm{Re}\,\hat E$, $\Delta_{\rm Im}=\mathrm{Im}\,\bar E-\mathrm{Im}\,\hat E$, $\bar E$ the window-averaged local energy, and $g(x)=\max(|x|-\delta,0)$. Within the tolerance region, $|\Delta_{\rm Re}|<\delta_{\rm Re}$ and $|\Delta_{\rm Im}|<\delta_{\rm Im}$, the guide term is exactly zero and Eq.~\eqref{eq:supp-psm-guide} reduces to the unguided PSM objective. The exact resonance pole is never used; only a coarse estimate with a much wider tolerance is needed. For $V_2=0.25$ we use guide weight $w=1$ and half-widths $\delta_{\rm Re}=0.5$, $\delta_{\rm Im}=0.1$; the useful restriction is on $\delta_{\rm Im}$, which keeps the optimizer away from the near-real quasi-eigenstate described above, without otherwise biasing the fit.

\begin{figure}[t]
 \centering
  \includegraphics[width=0.62\linewidth]{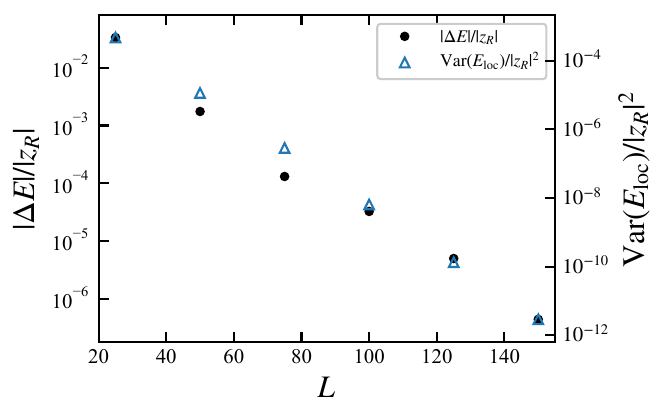}
 \caption{\label{fig:psm-broad-resonance}
 Guided PSM for the broad resonance $V_2=0.25$ ($\Gamma/E_R\approx0.97$): relative pole error (left axis) and the local-energy variance $\mathrm{Var}[E_{\rm loc}]/|z_R|^2$ used in the optimization (right axis) versus volume $L$.}
\end{figure}

With this guide, using volume scaling, the time window $t \in [0.125L,0.275L]$, and $L_\mathrm{init}=0.15L$, the relative pole error falls from $3.3\times10^{-2}$ at $L=25$ to $4.4\times10^{-7}$ at $L=150$. It tracks $\mathrm{Var}[E_{\rm loc}]$ over the same range (Fig.~\ref{fig:psm-broad-resonance}). Thus, the PSM keeps exponential finite-size convergence well beyond the narrow-resonance regime.

In contrast, a simple CAP strategy that tracks a single eigenvalue as $\eta$ is turned on is not always reliable for broad resonances. As $\Gamma$ grows, the resonance branch lies closer to nearby finite-box levels. The tracked trajectory can then pass through an avoided crossing before reaching the stationary point, so the calculation follows the wrong branch. For $V_2=0.25$, the tracked branch does not converge to the resonance pole over the volumes studied.

\subsection{Implementation details}
\label{sec:supp-psm-implementation}

\subsubsection{Calculations using classical algorithms}
For each volume, we first diagonalize the Hamiltonian restricted
to $L_{\mathrm{init}}$ and use the box eigenstate nearest a rough
target energy $E_{\mathrm{target}}$ as the initial guess. This
state is embedded in the larger volume and evolved with the full
Hermitian Hamiltonian. The target energy is used only to select
the initial box state. The optimization varies the normalized
real coefficients $c_n$ within $L_{\mathrm{init}}$. Using the JAX
library, we compute derivatives of the local-energy variance, and
the BFGS algorithm minimizes
\begin{equation}
 \mathcal{L}_{\mathrm{BFGS}}
 =
 \sqrt{\operatorname{Var}_t(E_{\mathrm{loc}})+\epsilon^2}
 -\epsilon,
 \qquad
 \epsilon=10^{-12}.
 \label{eq:supp-psm-bfgs-loss}
\end{equation}

We use the optimization windows $[15,25]$, $[15,45]$, and
$[0.7,1.3]\,\mathrm{MeV}^{-1}$ for the 1D two-body, 3D
two-body, and 1D three-body systems, respectively. Their upper
endpoints are chosen to precede the earliest estimated return
from the outer boundary, while the lower endpoints are 
representative choices that allow the PSM to produce low-variance plateaus in
$E_{\rm loc}(t)$.

To estimate the earliest return conservatively, we use the
maximum group velocity over the full lattice band, independently
of $E_{\mathrm{target}}$. For a nearest-neighbor lattice
direction,
\begin{equation}
 v_{\max}=2t_{\rm hop}a,
 \qquad
 t_{\mathrm{ret}}\simeq
 \frac{2(L-L_{\mathrm{init}})}{v_{\max}}.
 \label{eq:supp-psm-return-time}
\end{equation}
The factor of two accounts for propagation to the boundary and
back. In multiple dimensions, we use the fastest coordinate
direction.

For the 1D two-body system, $a=0.25$, $\mu=1$, and
$t_{\rm hop}=1/(2\mu a^2)=8$, giving $v_{\max}=4$ and
$t_{\mathrm{ret}}\simeq2(60-7)/4=26.5$. For the 3D two-body system,
$a=\mu=1$ and $t_{\rm hop}=1/2$ in each Cartesian direction, giving
$v_{\max}=1$ and $t_{\mathrm{ret}}\simeq2(60-7)/1=106$. For the
three-body system, $\mu_1=m/2$ and $\mu_2=2m/3$ give maximum
velocities of $82.942$ and $62.206\,\mathrm{MeV\,fm}$ along
$y_1$ and $y_2$, respectively. The faster $y_1$ direction
therefore gives
$t_{\mathrm{ret}}\simeq2(60-4)/82.942
=1.35034\,\mathrm{MeV}^{-1}$.

The pole is extracted from the survival amplitude inside the
optimization window. Table~\ref{tab:psm-figure-values} lists
the PSM calculations shown in Fig.~\ref{fig:surv_amps}, and
the reference poles are given in
Sec.~\ref{sec:supp-benchmarks}.
\subsubsection{Quantum rodeo/HVA calculation}
For the quantum test, $H_{\rm init}$ is the part of the physical Hamiltonian
inside $L_{\rm init}=7$, with hopping beyond that boundary removed.
The production Hamiltonian has radial extent $L=120$ and lattice
spacing $a=0.25$. All state preparation stays inside the compact
region. The finished state is embedded in the larger volume only when
the survival amplitude is measured.

\begin{table*}[t]
\caption{\label{tab:psm-figure-values} Figure-matched classical PSM
calculations. The entries give the production linear size, compact support,
lattice spacing, optimization window, and extracted pole. Lengths in the
three-body row are in fm, its time window is in $\mathrm{MeV}^{-1}$, and its
energy is in MeV.}
\begin{ruledtabular}
\begin{tabular}{lccccc}
System & $L$ & $L_{\rm init}$ & spacing & time window & $E_{\rm PSM}$ \\
\colrule
1D two-body & $60$ & $7$ & $0.25$ & $[15,25]$
& $1.081754052720-0.000536238798\,i$ \\
3D two-body & $60$ & $7$ & $1.0$ & $[15,45]$
& $1.3826301-0.0057088\,i$ \\
1D three-body & $60$ & $4$ & $1.0$ & $[0.7,1.3]$
& $8.1847589 - 1.0135397\,i$ \\
\end{tabular}
\end{ruledtabular}
\end{table*}

The input is a normalized real Gaussian, $\psi_n\propto e^{-r_n^2/2}$. A successful rodeo cycle applies
\begin{equation}
 M_k=\frac{1}{2}\left[
 e^{+i(H_{\rm init}-E_{\rm target})t_k}
 +e^{-i(H_{\rm init}-E_{\rm target})t_k}\right]
 =\cos[(H_{\rm init}-E_{\rm target})t_k],
 \label{eq:supp-quantum-rodeo-filter}
\end{equation}
followed by normalization. This operation is implemented by preparing one
ancilla in a superposition, controlling forward and backward real-time
evolution on its two branches, and postselecting the appropriate $X$-basis
outcome. A failed cycle requires restarting the state-preparation sequence.
The deterministic time samples are
\begin{equation}
 t_k=k^{1.25},\qquad k=1,\ldots,N_{\rm cycles},
 \label{eq:supp-quantum-rodeo-times}
\end{equation}
and we use the coarse target $E_{\rm target}=1.082$. This target is an input
to the filter and can be obtained from a coarse energy scan; the independent
lattice Schr\"odinger equation pole is not used. The cumulative postselection probabilities are
$0.1026$, $0.1022$, and $0.1022$ for $N_{\rm cycles}=6,8,10$, respectively.

For the HVA, split the compact Hamiltonian into its diagonal and hopping
parts,
\begin{equation}
 H_{\rm init}=H_D+H_K,\qquad
 H_D=\operatorname{diag}(H_{\rm init}),\qquad
 H_K=H_{\rm init}-H_D .
 \label{eq:supp-quantum-hva-split}
\end{equation}
At depth $p$, the compact variational circuit is
\begin{equation}
 U_{\rm HVA}(\boldsymbol\beta,\boldsymbol\gamma)
 =\prod_{\ell=1}^{p}
 e^{-i\beta_\ell H_K}e^{-i\gamma_\ell H_D}.
 \label{eq:supp-quantum-hva}
\end{equation}
Both factors are unitary. We evaluate the PSM loss at 20 times in
$t\in[15,60]$ with $\tau=0.5$. The real and imaginary parts of each survival
amplitude can be obtained on quantum hardware with an ancilla interferometric
measurement of controlled $e^{-iHt}$. We construct
\begin{equation}
E_{\rm loc}(t)
=
\frac{i}{\tau}
\log\!\left[
\frac{A(t+\tau)}{A(t)}
\right].
\end{equation}
For the calculations reported here, the sampled trajectory does not cross a
branch cut of the complex logarithm, so no explicit phase-unwrapping procedure
is required.

When the HVA depth is increased, the existing parameters are retained,
\begin{equation}
 \beta_k^{(p)}=\beta_k^{(p-1)},\qquad
 \gamma_k^{(p)}=\gamma_k^{(p-1)},\qquad k<p,
 \label{eq:supp-quantum-warm-start}
\end{equation}
and the newly appended layer is initialized uniformly in $[0,0.05]$.
For the ideal noiseless test in Table~\ref{tab:quantum_comp}, exact
statevector linear algebra emulates the unitary evolution, and
automatic differentiation with a strong-Wolfe LBFGS optimizer is used only
for classical convenience. Neither exact statevector access nor automatic
differentiation is required on a quantum computer. On hardware, the
same scalar loss can be reconstructed from measured survival amplitudes and
minimized with finite-difference or gradient-free classical optimization.
No CAP pole, lattice Schr\"odinger equation pole, or resonance eigenvector enters the rodeo filter, HVA,
or loss function; the lattice Schr\"odinger equation result is used only after extraction to form the
error columns in Table~\ref{tab:quantum_comp}.
This rodeo/HVA test is separate from the classical real-wavefunction
calculations in Table~\ref{tab:psm-figure-values}, which use the method
described above.

\subsubsection{Multiprecision refinement}

For Fig.~\ref{fig:psm-mp-comparison}, the half-line lattice has $n=0,\ldots,L$, with $L=25,50,\ldots,150$. At each $L$, we start from a new normalized real state
\begin{equation}
 |\psi\rangle=\sum_{n=0}^{L_{\mathrm{init}}}c_n|n\rangle,
 \qquad c_n\in\mathbb{R},\qquad
 L_{\mathrm{init}}=0.2L,
 \label{eq:supp-mp-psm-state}
\end{equation}
so $L_{\mathrm{init}}=5,10,\ldots,30$ for the plotted sequence. Each
calculation starts from the eigenstate of the compact Hamiltonian nearest the
pole-blind rough energy $E=1.04$.
The optimization and principal extraction window is
\begin{equation}
 t\in[t_{\min},t_{\max}]=[0.9L,1.1L],
 \qquad t_c=1.0L,
 \label{eq:supp-mp-psm-window}
\end{equation}
which follows the intermediate-time region as the box grows. 

First, 64-bit JAX and BFGS optimize the real coefficients $c_n$. With $A(t)=\langle\psi|e^{-iHt}|\psi\rangle$, define the
discrete local energy
\begin{equation}
 E_{\mathrm{loc}}(t_j)=\frac{i}{\Delta t}
 \log\!\left[\frac{A(t_j+\Delta t)}{A(t_j)}\right],\qquad
 \operatorname{Var}_t(E_{\mathrm{loc}})
 =\frac{1}{N_t}\sum_j|E_{\mathrm{loc}}(t_j)-\overline {E_{\mathrm{loc}}}|^2.
 \label{eq:supp-mp-psm-eloc}
\end{equation}
The float64 grid uses $\Delta t=0.05$, and BFGS minimizes
$\sqrt{\operatorname{Var}_t(E_{\mathrm{loc}})+\epsilon^2}-\epsilon$ with
$\epsilon=10^{-12}$. This stage is used only to find the correct minimum. We then switch the Hamiltonian,
state, amplitudes, and linear algebra to 50-decimal mpmath arithmetic. A
Gauss--Newton iteration minimizes the real and imaginary parts of the
centered $E_{\mathrm{loc}}$ values on the sparser collocation grid
$\Delta t=0.5$. We normalize after every update, remove one redundant
scale direction, and accept only line-search steps that reduce
the centered-residual norm.

Both the multiprecision refinement and the final evaluation apply the sparse
Hamiltonian without diagonalization. After scaling its conservative
spectral interval $[-4,5]$ to $[-1,1]$, we use
\begin{equation}
 e^{-iHt}=e^{-i\bar E t}\left[J_0(Rt)I
 +2\sum_{m=1}^{N_C}(-i)^mJ_m(Rt)T_m(\widetilde H)\right],
 \quad \widetilde H=\frac{H-\bar E}{R},
 \label{eq:supp-mp-psm-chebyshev}
\end{equation}
and generate $T_m(\widetilde H)|\psi\rangle$ by the three-term Chebyshev
recurrence; here $\bar E=0.5$ and $R=4.5$. The final survival amplitude is
sampled independently at
$\Delta t=0.1$ and fitted to
$\log A(t)=C-iE_{\mathrm{PSM}}t$. Depending on $L$, the retained orders are
$N_C=500$--$1200$; repeating the extraction at $N_C-80$, $N_C-40$, and $N_C$
checks truncation convergence. 
Neither the Gauss--Newton residual nor any survival-amplitude fit contains the
reference pole; it enters only after extraction through
$|E_{\mathrm{PSM}}-z_R|/|z_R|$.

As $L$ and $L_{\rm init}$ grow, the PSM pole approaches the exact lattice pole. Figure~\ref{fig:psm-mp-comparison} shows this convergence.

\end{document}